\documentclass[11pt,letterpaper]{article}
\pdfoutput=1

\usepackage[utf8x]{inputenc}%
\usepackage{tcolorbox,fancybox}%
\usepackage{physics}
\usepackage{ntheorem}

\usepackage[english]{babel}
\usepackage{amsmath,amssymb,amsfonts}
\numberwithin{equation}{section}
\usepackage{graphicx}
\usepackage{booktabs}
\usepackage{color,xcolor}
\usepackage[numbers,sort&compress]{natbib}

\usepackage{tikz}
\usepackage[compat=1.1.0]{tikz-feynman}
\usetikzlibrary{positioning}
\usetikzlibrary{decorations.text}
\usetikzlibrary{decorations.pathmorphing}

\usetikzlibrary{calc}
\usetikzlibrary{shapes.misc}
\tikzset{
        >=latex,
    photon/.style={decorate, decoration={snake}, draw=black, thick},
    fermionnoarrow/.style={draw=black, postaction={decorate}, thick},
    scalar/.style={draw=black, postaction={decorate}, decoration={markings,mark=at position .55 with {\arrow{>}}}, thick, dashed},
    scalarnoarrow/.style={draw=black, postaction={decorate},  thick, dashed},
    fermion/.style={draw=black, postaction={decorate},decoration={markings,mark=at position .55 with {\arrow{>}}}, thick},
    gluon/.style={decorate, draw=black, decoration={coil,amplitude=4pt, segment length=5pt}, thick},
    vertex/.style={draw,shape=circle,fill=black,minimum size=3pt,inner sep=0pt},
    fillvertex/.style={draw,shape=circle,fill=black,minimum size=5pt,inner sep=0pt},
    openvertex/.style={draw,shape=circle,minimum size=5pt,inner sep=0pt},
    blob/.style={draw=red,shape=circle,fill=red,minimum size=6pt,inner sep=0pt},
    redvertex/.style={draw=red,shape=circle,fill=red,minimum size=3pt,inner sep=0pt},
    cross/.style={cross out, draw=black,thick, minimum size=5pt, inner sep=0pt, outer sep=0pt}
}

\usepackage{physics}
\usepackage{orcidlink}
\usepackage{slashed}
\usepackage{mathrsfs}
\usepackage{enumerate}
\usepackage{mdframed}
\usepackage{url}
\usepackage[makeroom]{cancel}
\usepackage{geometry}

\usepackage{pifont}

\newtheorem*{thm-non}{Theorem}

\usepackage{hyperref} %Automatically links \label and \ref commands; Always load last
\hypersetup{
    colorlinks=true,       % false: boxed links; true: colored links
    linkcolor=red,          % color of internal links
    citecolor=blue,        % color of links to bibliography
    filecolor=magenta,      % color of file links
    urlcolor=purple           % color of external links
}
\usepackage[all]{hypcap} %Link navagates to top of figure instead of caption (below fig)

\def\beqn{\begin{eqnarray}}
\def\eeqn{\end{eqnarray}}
\def\beqs{\begin{subequations}}
\def\eeqs{\end{subequations}}
\def\beq{\begin{equation}}
\def\eeq{\end{equation}}
\def\ba{\begin{array}}
\def\ea{\end{array}}

\def\non{\nonumber\\}
\def\f{\frac}
\def\hf{\frac{1}{2}}
\def\[{\left[}
\def\]{\right]}
\def\({\left(}
\def\){\right)}
\newcommand\para{\paragraph{}}

\def\gG{\rm G}

\newcommand{\rep}[1]{\mathbf{#1}}
\newcommand{\repb}[1]{\mathbf{\overline{#1}}}

\def\Bc{\mathcal{B}}

\def\Dc{\mathcal{D}}
\def\Ec{\mathcal{E}}

\def\Gc{\mathcal{G}}
\def\Hc{\mathcal{H}}

\def\Lc{\mathcal{L}}

\def\Oc{\mathcal{O}}

\def\Qc{\mathcal{Q}}

\def\Tc{\mathcal{T}}
\def\Uc{\mathcal{U}}

\def\Wc{\mathcal{W}}
\def\Xc{\mathcal{X}}
\def\Yc{\mathcal{Y}}
\def\Zc{\mathcal{Z}}

\def\DG{\mathfrak{D}}  
  \def\eG{\mathfrak{e}}
  
  \def\gG{\mathfrak{g}}

  \def\nG{\mathfrak{n}}

  \def\uG{\mathfrak{u}}

\title{ 
 {\bf The non-topological $Z^\prime$ string in the 331 model \\ and its classical stability } \\
\author{\large Zhengyang Bian$^{1}$\,\orcidlink{0009-0000-6083-5518}, Ning Chen$^{2}$\,\orcidlink{0000-0002-0032-9012}, Mian Guo$^{3}$\,\orcidlink{0009-0002-7266-8467}, Zhanpeng Hou$^4$\,\orcidlink{0000-0002-6035-368X}, \\ Haoyang Ji$^5$\,\orcidlink{0009-0001-3013-9009}, Junyi Wei$^6$\,\orcidlink{0009-0006-4237-3456}, Zhuo Zhang$^7$\,\orcidlink{0009-0002-5415-6556}}
\date{\small \it
$^{1}\,^2\, ^3 \,^{5}\, ^6\, ^7$School of Physics, Nankai University, Tianjin, 300071, China \\
$^4$Institute of Particle Physics and Key Laboratory of Quark and Lepton Physics (MOE), \\ Central China Normal University, Wuhan, Hubei 430079, China \\
}
}

\begin{document}

\maketitle
\setlength{\parskip}{0.2ex}

\begin{abstract}
\bigskip
We study the classical stability of a non-topological $Z^\prime$ string in the minimal 331 model, which arises from the maximal symmetry breaking pattern of an ${\rm SU}(6)$ toy model. Two Higgs triplets are introduced according to the emergent global symmetries in the fermionic sector of the ${\rm SU}(6)$ toy model, which will achieve the sequential symmetry breaking of ${\rm SU}(3)_c\otimes {\rm SU}(3)_W \otimes {\rm U}(1)_X\to {\rm SU}(3)_c\otimes {\rm SU}(2)_W \otimes {\rm U}(1)_Y$. By analyzing small perturbations around the string background and solving the coupled Helmholtz equations numerically, we find that the string is stable only near the semilocal limit of $\vartheta_S \approx \frac{\pi}{2}$, even when Higgs self-couplings are tuned to minimize instabilities. This suggests that such non-topological strings are unlikely to exist in unified theories based on ${\rm SU}(N>5)$ Lie algebras.

\end{abstract}

\vspace{6.8cm}
{\emph{Emails:}\\  
$^{\,1}$\url{mronebian@mail.nankai.edu.cn}\\
$^{\,2}$\url{chenning_symmetry@nankai.edu.cn}\\
$^{\,3}$\url{guomian212025@mail.nankai.edu.cn}\\
$^{\,4}$\url{houzhanpeng@mails.ccnu.edu.cn}\\
$^{\,5}$\url{jihaoyang@mail.nankai.edu.cn}\\
$^{\, 6}$ \url{weijunyi@mail.nankai.edu.cn}\\
$^{\, 7}$ \url{zhuozhang@mail.nankai.edu.cn}\\
 }

\thispagestyle{empty}  
\newpage  
 
\setcounter{page}{1}  

\vspace{1.0cm}
\tableofcontents

%###################################################################
\vspace*{3mm}
\section{Introduction}
\label{section:intro}
%###################################################################
%
%

\para
Topological defects are ubiquitous in various new physics beyond the Standard Model (SM) when they undergo different symmetry-breaking stages according to the Kibble-Zurek mechanism~\cite{Kibble:1976sj,Zurek:1985qw}.
Among them, the $2+1$ dimensional vortices and $3+1$ dimensional strings originate from the discrete symmetry breaking pattern of $\Gc \to \Hc$ that has non-trivial first homotopy group of $\pi_1( \Gc / \Hc )\neq \mathbf{1}$.
The most well-known solution is the Abrikosov-Nielsen-Olesen (ANO) string~\cite{Abrikosov:1956sx,Nielsen:1973cs} in the Abelian Higgs model with the spontaneous breaking of ${\rm U }(1) \to \emptyset$.

\para
Decades ago, the Abelian string solutions were suggested to be embedded into the spontaneously broken non-Abelian theories, such as the electroweak (EW) sector of the SM.
Obviously, the spontaneous EW symmetry breaking has a trivial first homotopy group of $\pi_1({\rm SU}(2)_W \otimes {\rm U}(1)_Y / {\rm U}(1)_{\rm EM})=\mathbf{1}$.
There can be strings with ends, which are also known as the Nambu monopoles or dumbbells~\cite{Nambu:1977ag}.
There can also be strings without ends, which were proposed in Refs.~\cite{Vachaspati:1991dz,Hindmarsh:1991jq,Vachaspati:1992fi,Vachaspati:1992jk}, which are known as the $Z$ strings or the EW strings (cf. Ref.~\cite{Achucarro:1999it} for the review).
Different from the ANO strings, the stabilities of $Z$ strings are not topologically protected, while they depend on two parameters of the SM, namely, the Weinberg angle $\vartheta_W$ and the SM Higgs boson mass of $m_H$.
Unfortunately, a stable $Z$ string can only exist in the semilocal limit where the Weinberg angle approaches to $\vartheta_W\to \frac{\pi}{2}$ limit~\footnote{The semilocal limit of $\vartheta_W\to \frac{\pi}{2}$ in the SM corresponds to the large ${\rm U}(1)_Y$ coupling versus the vanishing non-Abelian ${\rm SU}(2)_W$ coupling. In such a limit, the local ${\rm SU}(2)_W$ invariance is reduced to a global $\widetilde {\rm SU}(2)$ symmetry. The corresponding first homotopy group tends to be non-trivial.} together with a light SM Higgs Higgs boson of $m_H< m_Z$.
Comparing with the experimental measurements~\cite{ATLAS:2012yve,CMS:2012qbp}, one cannot expect a stable $Z$ string in the SM.

\para
The 331 model was first proposed in Refs.~\cite{Lee:1977qs,Lee:1977tx}.
Such extended weak sector can naturally emerge as the subalgebra of the ${\rm SU}(8)$ grand unified theory, which is recently proposed as the minimal framework to embed three-generational SM quarks/leptons non-trivially~\cite{Barr:2008pn,Chen:2023qxi,Chen:2024cht,Chen:2024deo,Chen:2024yhb,Chen:2025ezv}.
As an extension to the SM Lie algebra, we are motivated to look for the non-topological string solutions and analyze their stabilities in the context of the minimal 331 model.
Through the detailed analysis, we wish to know if they can exist in unified theories with extended Lie algebras of ${\rm SU}(N>5)$.

\para
To analyze the classical stability, one applies small perturbations to all gauge fields and scalar fields in the string background.
One approach is to determine the stabilities through the signs of the energy variation due to the perturbations~\cite{James:1992zp,James:1992wb,Kanda:2022xrz,Kanda:2023yyz}.
In this work, we will determine the string stability by solving the eigenvalue equations from the stability matrix derived from the $Z^\prime$ string.
This approach was previously used for the stability analysis for the non-topological $Z$ string in Refs.~\cite{Goodband:1995rt,Goodband:1995he,Eto:2024xvc}.

\para
The rest of the paper is organized as follows.
In Sec.~\ref{section:setup}, we setup the 331 model from a one-generational ${\rm SU}(6)$ toy model.
The minimal fermionic sector enjoys a global non-anomalous $\widetilde {\rm SU}(2) \otimes \widetilde {\rm U}(1)_T$ symmetry, and two Higgs triplet fields are assumed according to the global symmetries.
In Sec.~\ref{section:string}, we obtain the classical $Z^\prime$ string solution in the 331 model.
With the numerical solutions of the $Z^\prime$ string profile functions, we carry out the detailed analysis of the string stability in Sec.~\ref{section:stability_331}, where we employ the small perturbations to both the Higgs fields and the gauge fields.
Based on the stability matrix to the perturbed Fourier modes, we find the stable regions numerically in Sec.~\ref{section:numerical}, and they point to the semilocal limit of the 331 mixing angle of $\vartheta_S\to \f{\pi}{2}$ (to be defined in Eq.~\eqref{eq:Salam_angle} below).
We summarize our results and comment on future searches in Sec.~\ref{section:conclusion}.
In Appendix~\ref{section:SDeq}, we derive the self-dual equations for the ${\rm SU}(N) \otimes  {\rm U}(1)_X\to {\rm SU}(N-1) \otimes  {\rm U}(1)_{X^\prime}$ breaking pattern.
The critical point that saturate the Bogmol'nyi bound is obtained as well.

%###################################################################
\vspace*{3mm}
\section{The 331 model from an ${\rm SU}(6)$ toy model}
\label{section:setup}
%###################################################################

\para
In this section, we setup the 331 model as an effective theory from the maximal symmetry breaking pattern of a one-generational ${\rm SU}(6)$ toy model~\cite{Chen:2021haa,Chen:2021zwn}.

\subsection{The fermion sector}

\para
The maximally ${\rm SU}(6)$ breaking pattern is
\beqn\label{eq:SU6_pattern}
&& {\rm SU }(6) \xrightarrow{ v_U } {\Gc }_{331}  \xrightarrow{ v_{331}}  {\Gc }_{\rm SM}  \,,\non
&& {\Gc }_{331} = {\rm SU }(3)_c \otimes  {\rm SU }(3)_W \otimes  {\rm U}(1)_X \,,\quad  {\Gc }_{\rm SM}= {\rm SU }(3)_c \otimes  {\rm SU }(2 )_W \otimes  {\rm U}(1)_Y \,,
\eeqn
where the GUT-scale symmetry breaking is achieved by the following Higgs VEV of an ${ \rm SU}(6)$ adjoint Higgs field of $\rep{35_H}$
\beqn
&& \langle \rep{35_H} \rangle = \frac{1}{2 \sqrt{3} } {\rm diag} (  - \mathbb{I}_3 \,, + \mathbb{I}_3  ) v_U \,.
\eeqn
The ${\rm U}(1)_X$ charge operator for the $\rep{6} \in {\rm SU }(6)$ and ${\rm U}(1)_Y$ charge operator for the $\rep{3}_W/\repb{3}_W \in {\rm SU}(3)_W$ are defined by
\beqs
\beqn
X( \rep{6} ) &\equiv& {\rm diag} (  -\frac{1}{3} \mathbb{I}_3 \,, +\frac{1}{3} \mathbb{I}_3  ) \,,\label{eq:Xcharge_331}\\[2mm]
 Y( \rep{3}_W )&\equiv& {\rm diag} (  ( \frac{1}{6} + \Xc ) \mathbb{I}_2 \,, -\frac{1}{3} + \Xc ) \,, \label{eq:Ycharge_331} \\[2mm]
  Y( \repb{3}_W )&\equiv& {\rm diag} (  ( -\frac{1}{6} + \bar \Xc ) \mathbb{I}_2 \,, +\frac{1}{3} + \bar \Xc) \,, \label{eq:Ycharge_331}
\eeqn
\eeqs
where $\Xc/ \bar \Xc$ represent the ${\rm U}(1)_X$ charges of the ${\rm SU}(3)_W$ fundamental and anti-fundamental representations, respectively.
The electric charge operator of the ${\rm SU }(3)_W$ fundamental and antifundamental representations are expressed as a $3\times 3$ diagonal matrix
\beqn\label{eq:Qcharge_331}
Q ( \rep{3}_W )&\equiv& T_{{\rm SU }(3)}^3 + Y( \rep{3}_W ) = {\rm diag} ( \frac{2}{3}+ \Xc \,,- \frac{1}{3}+ \Xc \,, -\frac{1}{3}+ \Xc )\,, \non
Q ( \repb{3}_W )&\equiv& -T_{{\rm SU }(3)}^3 + Y( \repb{3}_W ) = {\rm diag} (- \frac{2}{3}+ \bar \Xc \,, + \frac{1}{3}+ \bar \Xc \,, +\frac{1}{3}+ \bar \Xc )\,, \non
{\rm with}&~&  T_{{\rm SU }(3)}^3= \frac{1}{2} {\rm diag} (1\,,-1 \,,0 ) \,.
\eeqn 

\begin{table}[htp]
\begin{center}
\begin{tabular}{c|c|c}
\hline \hline
   ${\rm SU}(6)\,, \[  \widetilde {\rm SU}(2)_F \otimes \widetilde {\rm U}(1)_T \]$   &  ${\Gc }_{331}\,, \[  \widetilde {\rm SU}(2)_F \otimes \widetilde {\rm U}(1)_{T^\prime} \]$  & ${\Gc }_{\rm SM}$  \\
\hline \hline
  $\repb{6_F}^1 \,, ( \rep{2}\,,  -2t)$ & $(\repb{ 3} \,, \mathbf{1} \,, + \frac{1}{3})_{ \mathbf{F} }^{1 } \,, ( \rep{2}\,,  -t) ~:~ (\Dc_R^{1 })^c$    &  $( \repb{ 3} \,, \mathbf{1} \,, + \frac{1}{3})_{ \mathbf{F} }^{1 }~:~ {d_R}^c  $ \\[1mm]   
    & $(\rep{1} \,, \repb{3} \,,  -\frac{1}{3})_{ \mathbf{F} }^{1}\,, ( \rep{2}\,,  -3t) ~:~ \Lc_L^{1 }$  &  $( \rep{1} \,, \repb{ 2}  \,, -\frac{1}{2} )_{ \mathbf{F} }^{1 }~:~  \ell_L = ( e_L \,, -\nu_L)^T $\\[1mm] 
    &  & $( \mathbf{1} \,, \mathbf{ 1}  \,, 0 )_{ \mathbf{F} }^{1 }~:~ \check \nG_{L}^1$  \\[1mm] \hline
     %%%%%%%%%%%%%%%%%%%%%%%%%%%%%%%%%%%%%%%%%%%%%
 $\repb{6_F}^2 \,, ( \rep{2}\,,  -2t)$  & $(\repb{ 3} \,, \mathbf{1} \,, + \frac{1}{3})_{ \mathbf{F} }^{2} \,, ( \rep{2}\,,  -t)~:~ (\Dc_R^{2})^c $   &  $( \repb{3}\,, \mathbf{1} \,, + \frac{1}{3})_{ \mathbf{F} }^{2 }~:~ { \DG_R}^c$ \\[1mm]
    &  $(\mathbf{1} \,, \repb{ 3} \,,  -\frac{1}{3})_{ \mathbf{F} }^{2 } \,, ( \rep{2}\,,  -3t)~:~ \Lc_L^{2 }$ &  $( \rep{1} \,, \repb{ 2}  \,, -\frac{1}{2} )_{ \mathbf{F} }^{2 } ~:~ ( \eG_L \,, - \nG_L )^T $ \\[1mm] 
   &   &  $( \mathbf{1} \,, \mathbf{1}  \,, 0 )_{ \mathbf{F} }^{2 }~:~  \check \nG_L^2 $  \\[1mm] \hline
    %%%%%%%%%%%%%%%%%%%%%%%%%%%%%%%%%%%%%%%%%%%%%
    $\rep{15_F} \,, ( \rep{1}\,,  +t)$ & $( \repb{3} \,, \mathbf{1} \,, - \frac{2}{3})_{ \mathbf{F}} \,, ( \rep{1}\,,  -t) ~:~ {u_R}^c$  & $( \repb{3} \,, \mathbf{1} \,, - \frac{2}{3})_{ \mathbf{F} }~:~   {u_R}^c$ \\[1mm] 
    & $( \mathbf{1}\,, \repb{3} \,, + \frac{2}{3})_{ \mathbf{F}}\,, ( \rep{1}\,,  +3t) ~:~ ( \Ec_R)^c $  & $( \mathbf{1} \,, \repb{ 2}  \,, +\frac{1}{2})_{ \mathbf{F} }~:~ ( {\nG_R}^{c} \,, {\eG_R}^{c})^T $ \\[1mm] 
  &    & $( \mathbf{1} \,, \mathbf{1}  \,, +1)_{ \mathbf{F} }~:~ { e_R}^c $ \\[1mm]  
  & $(\mathbf{ 3} \,, \mathbf{3} \,, 0)_{ \mathbf{F} }\,, ( \rep{1}\,,  +t )~:~ \Qc_L$  &    $(\mathbf{ 3} \,, \mathbf{2} \,, +\frac{1}{6} )_{ \mathbf{F} }~:~  q_L=( u_L\,, d_L)^T $ \\[1mm]  
  &   &   $(\mathbf{ 3} \,, \mathbf{1} \,, -\frac{1}{3} )_{ \mathbf{F} }~:~  \DG_L$ \\[1mm] \hline
 %%%%%%%%%%%%%%%%%%%%%%%%%%%%%%%%%%%%%%%%%%%%%
\hline
\end{tabular}
\end{center}
\caption{
The ${\rm SU}(6)$ fermion representations under the ${\Gc }_{331}$ and the ${\Gc }_{\rm SM}$.
}
\label{tab:SU6_331_ferm}
\end{table}%

\para
The minimal anomaly-free ${\rm SU}(6)$ model contains the following left-handed fermions of 
\beqn
\{ f_L \}_{ \rm SU(6) } &=& \repb{6_F}^\omega \oplus \rep{15_F} \,,\quad \omega =1\,,2 \,.
\eeqn
The fermion sector enjoys a non-anomalous global symmetry of 
\beqn\label{eq:SU6_flavor}
\widetilde {\cal G}_{\rm flavor}&=& \widetilde {\rm SU}(2)_F \otimes \widetilde {\rm U}(1)_{T} \,,
\eeqn
according to Ref.~\cite{Dimopoulos:1980hn}.
Under the symmetry breaking pattern in Eq.~\eqref{eq:SU6_pattern} and the charge quantization given in Eqs.~\eqref{eq:Xcharge_331}, \eqref{eq:Ycharge_331}, and \eqref{eq:Qcharge_331}, we summarize the ${\rm SU}(6)$ fermions and their names in Tab.~\ref{tab:SU6_331_ferm}.
For the SM fermions marked by solid underlines, we name them by the first-generational SM fermions.
The global $ \widetilde {\rm U}(1)_{T}$ symmetry will become the global $ \widetilde {\rm U}(1)_{T^\prime}$ symmetry under the $\gG_{331}$ and the global $ \widetilde {\rm U}(1)_{B-L}$ symmetry under the $\gG_{\rm SM}$, and they are defined by~\cite{Chen:2023qxi}
\beqn\label{eq:BmL_def}
&& \Tc^\prime \equiv \Tc + 3t \, \Xc \,, \quad \Bc - \Lc \equiv \Tc^\prime\,.
\eeqn

\subsection{The Higgs sector}

\para
The most general Yukawa couplings that are invariant under the gauge symmetry are expressed as
\beqn\label{eq:SU6_Yukawa}
-\Lc_Y&=& Y_\Dc\,  \repb{6_F}^\omega \rep{15_F}  \repb{6_H}_{\,,\omega } +  Y_\Uc\, \rep{15_F} \rep{15_F}  \rep{15_H}   + H.c. \,.
\eeqn
The global $\widetilde {\rm U}(1)_T$ charges of the Higgs fields are given by
\beqn
&& \Tc( \repb{6_H}_{\,,\omega } )= +t \,, \quad  \Tc(  \rep{15_H} ) = -2t \,,
\eeqn
from the charge assignments in Tab.~\ref{tab:SU6_331_ferm} and Eq.~\eqref{eq:SU6_Yukawa}.
According to the symmetry breaking pattern in Eq.~\eqref{eq:SU6_pattern}, the 331 model consists of two ${\rm SU}(3)_W$ anti-fundamentals of $\Phi_{ \repb{3}\,,\omega } \equiv ( \rep{1}\,, \repb{3} \,, - \frac{1}{3})_{ \mathbf{H}\,, \omega } \subset \repb{6}_{ \mathbf{H} \,, \omega }$ (with $\omega=1\,,2$) and $\Phi_{ \repb{3} }^\prime \equiv ( \rep{1}\,, \repb{3} \,, +\frac{2}{3})_{ \mathbf{H}} \subset \rep{15_H}$ after the GUT-scale symmetry breaking as follows
\beqn\label{eq:Higgs_Br01}
&& \repb{6_H}_{\,,\omega } = ( \repb{3}\,,\rep{1}\,, +\frac{1}{3} )_{ \mathbf{H}\,, \omega } \oplus \overbrace{ ( \rep{1}\,,\repb{3}\,, -\frac{1}{3} )_{ \mathbf{H} \,, \omega }}^{ \rep{\Phi}_{ \repb{3}\,, \omega } } \,,\non
&& \rep{15_H} = ( \repb{3}\,,\rep{1}\,, -\frac{2}{3} )_{ \mathbf{H}} \oplus \overbrace{ ( \rep{1}\,,\repb{3}\,, +\frac{2}{3} )_{ \mathbf{H}} }^{ \rep{\Phi}_{\repb{3}}^\prime } \oplus ( \rep{3}\,,\rep{3}\,, 0 )_{ \mathbf{H}} \,, 
\eeqn
for the symmetry breaking pattern in Eq.~\eqref{eq:SU6_pattern}.
Two $( \rep{ 1} \,, \repb{3} \,, -\frac{1}{3})_{ \mathbf{H}\,, \omega } \subset \repb{6_H}_{\,,\omega}$ contain SM-singlet directions after the GUT-scale symmetry breaking in Eq.~\eqref{eq:SU6_pattern}, and they are also $\widetilde { {\rm U}}(1)_{T^\prime}$-neutral according to Eq.~\eqref{eq:BmL_def}. 
Meanwhile, the $( \rep{ 1} \,, \repb{2} \,, +\frac{1}{2} )_{ \mathbf{H}} \subset ( \rep{ 1} \,, \repb{3} \,, +\frac{2}{3})_{ \mathbf{H}} \subset \rep{15_H}$ can only develop VEV to trigger the spontaneous EWSB of ${\rm SU}(2)_W \otimes {\rm U}(1)_Y \to {\rm U}(1)_{\rm EM}$.
All ${\rm SU}(3)_c$ colored components of $(\repb{6_H}_{\,, \omega } \,, \rep{15_H})$ are assumed to obtain heavy masses of $\Lambda_{\rm GUT}$.
The residual massless Higgs fields transforming under the ${\rm SU}(3)_W \otimes {\rm U}(1)_X$ symmetry form the following Higgs potential
%
%
%\beqs
\beqn\label{eq:331_potential}
%V_{\rm tot}&=& V( \Phi_{ \repb{3}\,, \lambda } ) + V( \Phi_{ \repb{3}}^\prime ) + V( \Phi_{ \repb{3}\,, \lambda}\,,\Phi_{ \repb{3}}^\prime )   \,,\\[2mm]
%%%%%%%%%%%%%%%%%%%%%%%%%%%%%%%%%%%%%%%%%%%%%
V( \Phi_{ \repb{3}\,, \omega } )&=& \f{ \lambda_1 }{2} \( | \Phi_{ \repb{3}\,, 1 } |^2 - \hf V_1^2  \)^2 + \f{ \lambda_2}{2 } \( | \Phi_{ \repb{3}\,, 2 } |^2 - \hf V_2^2  \)^2 + \f{ \lambda_3 }{2 } \( | \Phi_{ \repb{3}\,, 1 }  |^2 + |  \Phi_{ \repb{3}\,, 2 } |^2  - \hf v_{331}^2   \)^2 \non
 && + \lambda_4 \( | \Phi_{ \repb{3}\,, 1 } |^2 |  \Phi_{ \repb{3}\,, 2 } |^2   - | \Phi_{ \repb{3} \,, 1}^\dag \Phi_{  \repb{3} \,, 2 } |^2  \) +  \lambda_5 \Big|   \Phi_{ \repb{3}\,, 1 }^\dag \Phi_{ \repb{3}\,, 2 }  - \hf V_1 V_2  \Big|^2  \,, 
\eeqn
%\eeqs
%
%
where $V_1^2 + V_2^2 = v_{331}^2$, and all parameters are assumed to be real.

\para
The Higgs fields responsible for the ${\rm SU}( 3 )_W \otimes {\rm U}(1)_X \to {\rm SU}(2)_W \otimes {\rm U}(1)_Y$ breaking are explicitly expressed as follows
\beqn\label{eq:331_Higgs}
&&  \Phi_{ \repb{3}\,, \omega }  = \frac{1}{ \sqrt{2}} \left( \ba{c}  \sqrt{2} \pi_{ W^\prime \,,\omega }^- \\    \sqrt{2} \pi_{ N \,, \omega }^0   \\    \sigma_\omega - i\pi^0_{Z^\prime\,, \omega  } \ea  \right) \,,
\eeqn
where the electric charges are given according to Eq.~\eqref{eq:Qcharge_331}.
According to the symmetry breaking pattern, we denote the Higgs VEVs of~\footnote{Throughout the paper, we use the short-handed notations of $(s_\vartheta\,, c_\vartheta\,, t_\vartheta)\equiv ( \sin\vartheta\,, \cos\vartheta\,, \tan\vartheta)$ for all mixing angles.}
\beqn\label{eq:331_VEV01}
&& \langle \sigma_1 \rangle= V_1 =  v_{331} c_{\tilde \beta} \,,\quad \langle \sigma_2 \rangle= V_2 =  v_{331} s_{\tilde \beta}   \,,
\eeqn
with 
\beqn
t_{\tilde \beta }&\equiv& \frac{V_2}{ V_1}
\eeqn
parametrizing the ratio between two $331$ symmetry breaking VEVs.
The Yukawa coupling term in Eq.~\eqref{eq:SU6_Yukawa} from two Higgs triplet VEVs lead to the following mass terms and mixing
\beqn
&& Y_\Dc\,  \repb{6_F}^\omega \rep{15_F}  \repb{6_H}_{\,,\omega } + H.c. \non
&\supset& Y_\Dc \, \Big[ (  \repb{3}\,, \rep{1}\,, + \frac{1}{3} )_{ \mathbf{F}  }^\omega  \otimes  ( \rep{3 }\,,\rep{3}\,,  0 )_{ \mathbf{F}  } \oplus  ( \rep{1}\,,\repb{3}\,, -\frac{1}{3} )_{ \mathbf{F}  }^\omega \otimes ( \rep{1}\,,\repb{3}\,, + \frac{ 2 }{3} )_{ \mathbf{F}  }  \Big] \otimes \langle ( \rep{1}\,,\repb{3}\,, -\frac{1}{3} )_{ \mathbf{H} \,, \omega } \rangle \non
&\Rightarrow& \frac{ Y_\Dc }{ \sqrt{2} } \[  ( e_L {\eG_R}^c + \nu_L {\nG_R}^c  + \DG_L {d_R}^c ) V_1 + ( \eG_L {\eG_R}^c + \nG_L {\nG_R}^c  + \DG_L {\DG_R}^c ) V_2  + H.c. \]\,. 
\eeqn
Apparently, the $(\eG\,, \nG\,, \DG)$ obtain the vectorlike fermion masses at this stage.

\subsection{The gauge sector}

\para
We express the ${\rm SU}(3)_W \otimes {\rm U}(1)_X$ covariant derivatives for the ${\rm SU}(3)_W $ anti-fundamental field $\Psi_{ \repb{3}} \equiv ( \repb{3} \,, \bar \Xc)$ as follows
%
%
%\beqs\label{eqs:331_covariant}
\beqn
%i D_\mu \Psi_{ \rep{3}} &\equiv& ( i \partial_\mu \mathbb{I}_3 +   g_{3W} W_\mu^{ I}  T_{ {\rm SU}(3) }^{ I} +  g_{X } \Xc  \mathbb{I}_3  X_{ \mu} ) \cdot  \Psi_{ \rep{3}} \,,\label{eq:331_covariant_fund} \\[2mm]
%%%%%%%%%%%%%%%%%%%%%%%%%%%%%%%%%%%%%%%%%%%%%
i D_\mu \Psi_{ \repb{3}}  &\equiv& \(  i \partial_\mu \mathbb{I}_3    - g_{3W} W_\mu^{ I} ( T_{ {\rm SU}(3) }^{ I} )^T + g_{X }  \bar \Xc  \mathbb{I}_3  X_{ \mu} \) \cdot \Psi_{\repb{3}} \,, \label{eq:331_covariant_antifund}
\eeqn
%\eeqs
%
%
where the ${\rm SU}(3)_W$ generators are normalized as ${\rm Tr}\(  T_{ {\rm SU}(3) }^{ I} T_{ {\rm SU}(3) }^{ J }  \)= \hf \delta^{ I   J}$.
The gauge fields from Eq.~\eqref{eq:331_covariant_antifund} can be expressed in terms of a $3\times 3$ matrix 
\beqn\label{eq:331_connection_gauge}
&& - g_{3W} W^{ I}_\mu  ( T_{ {\rm SU}(3) }^{ I} )^T + g_{X} \bar \Xc  \mathbb{I}_3  X_{ \mu}  \non
&=&  - \frac{g_{3W}}{ \sqrt{2} }\left( 
\ba{ccc}  
    \f{1}{\sqrt{2} } W_\mu^3  &   W_\mu^-  &  0 \\ 
 W_\mu^+  &  -  \f{1}{\sqrt{2} } W_\mu^3    &  0  \\ 
 0  &  0 &  0   \\  \ea  \right)  -  \frac{g_{3W}}{ \sqrt{2} }\left( 
\ba{ccc}  
     0  &  0  &   W_\mu^{\prime\, -}   \\ 
 0 &  0   &  \bar N_\mu  \\ 
 W_\mu^{\prime\, +}  & N_\mu  &  0   \\  \ea  \right)  \non
 &-& \frac{ g_{3W} }{2 \sqrt{3} } {\rm diag}  \(    ( W_\mu^8  - 6 t_{ \vartheta_S } \bar \Xc  X_{ \mu} ) \mathbb{I}_2 \,,  - 2  W_\mu^8  - 6 t_{ \vartheta_S } \bar \Xc X_{ \mu}  \)  \,,
\eeqn
with the 331 mixing angle $\vartheta_S$ defined by
\beqn\label{eq:Salam_angle}
t_{ \vartheta_S }&\equiv&  \frac{g_{X } }{ \sqrt{3} g_{3W } }\,.
\eeqn

\para
With the Higgs VEVs given in Eq.~\eqref{eq:331_VEV01}, the Higgs kinematic terms lead to the following mass terms for the gauge bosons
\beqn
 | D_\mu \Phi_{ \repb{3}\,, \lambda } |^2 &=& | ( \partial_\mu \mathbb{I}_3 + i g_{3W}  W_\mu^I ( T_{ {\rm SU}(3) }^I )^T - i g_X ( -\frac{1}{3} ) \mathbb{I}_3  X_\mu  ) \Phi_{ \repb{3}\,, \lambda }  |^2 \non
&\supset& g_{3W}^2  \Big|   \frac{ 1 }{2} \left( \ba{ccc}  \frac{1}{ \sqrt{3} } (W_\mu^8  + 2 t_{ \vartheta_S } X_\mu )& 0  & \sqrt{2} W_\mu^{\prime\, -}   \\ 
                                                0 &   \frac{1}{ \sqrt{3} } (W_\mu^8 + 2 t_{ \vartheta_S } X_\mu )  &  \sqrt{2} \bar N_\mu   \\
                                               \sqrt{2} W_\mu^{\prime\, +} &   \sqrt{2}  N_\mu   &   \frac{ 2 }{ \sqrt{3} }( - W_\mu^8 +   t_{ \vartheta_S } X_\mu ) \\    \ea  \right)  \non
                                               &&  \cdot  \frac{1}{ \sqrt{2} } \left( \ba{c}  0 \\   0   \\    V_\omega  \ea  \right)  \Big|^2 \non
&\supset& \frac{1}{ 4} g_{3W}^2 v_{ 331 }^2 \Big[  ( W_\mu^{ \prime\, +} W^{\prime \, - \, \mu } + N_\mu \bar N^\mu )  + \frac{2 }{3 } ( W_\mu^8 -t_{ \vartheta_S } X_\mu )^2 \Big]\,.
\eeqn
The massive gauge bosons of $(W_\mu^{\prime\, \pm} \,, N_\mu \,, \bar N_\mu)$ are due to the following off-diagonal components
\beqn
&& W_\mu^{ \prime\, \pm} \equiv \frac{ 1}{ \sqrt{2} } ( W_\mu^4 \mp i W_\mu^5) \,,\quad N_\mu \equiv  \frac{ 1}{ \sqrt{2} } ( W_\mu^6 - i W_\mu^7) \,, \quad  \bar N_\mu  \equiv  \frac{ 1}{ \sqrt{2} } ( W_\mu^6 + i W_\mu^7) \,,
\eeqn
and they carry the ${\rm U}(1)_Y$ charges of 
\beqn\label{eq:331Y_offGB}
&& \Yc(W_\mu^{\prime\, +})=\Yc( N_\mu) = + \f{1}{2} \,, \quad  \Yc(W_\mu^{\prime\, -})=\Yc( \bar N_\mu) = - \f{1}{2} \,.
\eeqn
The massive gauge boson $Z_\mu^\prime$ and the massless ${\rm U}(1)_Y$ gauge boson $B_\mu$ are related by the mixing angle in Eq.~\eqref{eq:Salam_angle} as follows
\beqn\label{eq:331_gauge_trans}
&& Z_\mu^\prime \equiv c_{\vartheta_S} W_\mu^8 - s_{  \vartheta_S } X_\mu \,, \non
&& B_\mu \equiv  s_{  \vartheta_S } W_\mu^8 + c_{  \vartheta_S } X_\mu \,.
\eeqn
By matching the ${\rm SU}(3)_W \otimes {\rm U}(1)_X$ gauge couplings of $( \alpha_{3W} \,, \alpha_X )$ with the EW gauge couplings of $( \alpha_{2W} \,, \alpha_Y )$ as follows
\beqn\label{eq:331_coupMatch}
&& \alpha_{2W}^{- 1} (v_{331} ) = \alpha_{3W}^{-1}  (v_{331} ) \,, \quad \alpha_{ Y}^{- 1} (v_{331} ) = \frac{1}{3} \alpha_{3W}^{-1}  (v_{331} ) + \alpha_{ X}^{-1}  (v_{331} ) \,, \non
&& \frac{1}{3} \alpha_{3W}^{-1} = \alpha_Y^{-1} s_{\vartheta_S }^2 \,,~ \alpha_{X }^{-1} = \alpha_Y^{-1} c_{ \vartheta_S }^2 \,,
\eeqn
we find the following matching relation
\beqn\label{eq:331_angleMatch}
&& s_{ \vartheta_S }= \frac{1 }{ \sqrt{3}} t_{ \vartheta_W }
\eeqn
between the ${\rm SU}(3)_W \otimes {\rm U}(1)_X$ mixing angle of $\vartheta_S$ and the EW Weinberg angle of $\vartheta_W$.
%If one takes the limit of $\vartheta_S \to \f{ \pi}{2}$, we have $Z_\mu^\prime \approx -X_\mu$ according to Eq.~\eqref{eq:331_gauge_trans} and $\vartheta_W \approx \frac{\pi}{3}$ according to the matching relation in Eq.~\eqref{eq:331_angleMatch}.
For our later convenience, we define a gauge coupling associated to the $Z^\prime_\mu$ as follows
\beqn\label{eq:331_ZpCoup}
&& g_{Z^\prime } \equiv \sqrt{ 3 g_{3W }^2 + g_X^2 } = \frac{ g_Y  }{ s_{\vartheta_S } c_{\vartheta_S } } \,, \quad  g_{ 3 W} = \frac{1 }{ \sqrt{3} } g_{Z^\prime } c_{ \vartheta_S } \,, \quad g_X = g_{Z^\prime } s_{ \vartheta_S } \,.
\eeqn
Thus, we find the mass squared of all massive gauge bosons to be
\beqn\label{eq:331_GBmass}
&& m_{W^{\prime\, \pm  }}^2 = m_{N \,, \bar N }^2 = \frac{ g_{ Y }^2  }{ 12 s_{ \vartheta_S }^2  }  v_{ 331}^2 = \frac{ 1  }{ 12   }   g_{Z^\prime }^2 c_{ \vartheta_S }^2 v_{ 331}^2  \,, \non
&&  m_{ Z^\prime }^2 = \frac{ g_{Y}^2 }{9  s_{ \vartheta_S }^2 c_{ \vartheta_S }^2  }  v_{ 331}^2 = \f{1}{9} g_{Z^\prime }^2 v_{ 331}^2 \,.
\eeqn

\para 
In terms of the 331 mass eigenstates, the covariant derivative for the ${\rm SU}(3)_W$ anti-fundamental representation reads
\beqn\label{eq:331_connection_antifund}
&& - \frac{g_{3W}}{ \sqrt{2} } \left( 
\ba{ccc}  
 \f{1}{ \sqrt{2} } W_\mu^3  & W_\mu^- & 0  \\ 
W_\mu^+  &  -  \f{1}{ \sqrt{2} } W_\mu^3 & 0 \\ 
0 &  0  &  0  \\  \ea  \right)  + g_Y {\rm diag}\,\Big( (-\frac{1}{6} + \bar \Xc ) \mathbb{I}_{2 } \,,  \frac{1}{3} + \bar \Xc \Big) B_\mu   \non
&& - \underline{ \frac{g_{3W}}{ \sqrt{2} } \left( 
\ba{ccc}  
 0  & 0  & W_\mu^{\prime\,-}  \\ 
0  &  0 & \bar N_\mu  \\ 
W_\mu^{\prime\, +}  & N_\mu  &  0  \\  \ea  \right) }  + \underline{ g_{Z^\prime }  {\rm diag} \Big( \[ - \frac{1}{6} +  ( \frac{1}{6} - \bar \Xc ) s_{\vartheta_S }^2  \] \mathbb{I}_{2  }  \,, \frac{1}{3} - (\frac{1}{3} + \bar  \Xc ) s_{\vartheta_S }^2  \Big) Z_\mu^\prime } \,,
\eeqn
where we highlight the massive gauge bosons at this symmetry breaking stage with underlines.

%###################################################################
\vspace*{3mm}
\section{The $Z^\prime$ string solution in the 331 model}
\label{section:string}
%###################################################################

\subsection{The $Z^\prime$ string profile}
\label{section:string_ansatz}

\para
In this section, we look for the $Z^\prime$ string solution in the 331 model.
The EW string solution was first proposed in Refs.~\cite{Vachaspati:1992fi,Vachaspati:1992jk}.
The $Z^\prime$ string solution with the unit winding number takes the following profile functions in the two-dimensional polar coordinates of $(r\,, \varphi)$
\beqn\label{eq:331_string_form}
&& \vec Z_{\rm ANO}^\prime = - \frac{ \bar \zeta ( r ) }{r } \vec e_\varphi  \,,\quad \Phi_{ \repb{ 3 } \,, \omega }  =  \left( 
\ba{c}  
 0    \\ 
 0 \\ 
  \phi_{ {\rm ANO} \,, \omega }  \\  \ea  \right)\,,\quad  \phi_{ {\rm ANO} \,, \omega }  =  \frac{ V_\omega }{ \sqrt{2} } \bar f_\omega ( r ) e^{i  \varphi }   \,,\quad \omega=(1\,, 2) \,,
\eeqn
while all other components of gauge fields are vanishing~\footnote{Our gauge field profile is similar to $Z$ string profile in Refs.~\cite{Vachaspati:1992fi,Kanda:2022xrz}, while differs from the $Z$ string profile in Refs.~\cite{Goodband:1995he,Eto:2024xvc} by a factor $\propto \f{ 2 }{g_{Z}}$. It turns out the profile function in Eq.~\eqref{eq:331_string_form} is convenient to obtain the boundary conditions in the 331 model, as well as in the generic ${\rm SU}(N) \otimes {\rm U}(1)$ models.}.
The $Z^\prime$ magnetic flux is quantized as
\beqn\label{eq:331_string_flux}
\Phi_{Z^\prime}&=& \int d^2 x \,  \partial_{ [1 } Z_{ {\rm ANO}\, 2] }^\prime= \f{ 6 \pi }{ g_{Z^\prime}} \,,
\eeqn
with the boundary condition for the $\bar \zeta(r)$ to be given in Eq.~\eqref{eqs:331string_boundary}.
We denote the covariant derivatives to anti-fundamental fields under the $Z^\prime$ string background as follows
%
%
%\beqs
\beqn
%d_m \Psi_{ \rep{3} }&\equiv& \Big\{ \partial_m \mathbb{I}_3  - i g_{Z^\prime } {\rm diag} \Big( (  \frac{1}{6} -  ( \frac{1}{6} +  \Xc ) s_{\vartheta_S }^2  ) \mathbb{I}_{2  }  \,, - \frac{1}{3} + (\frac{1}{3} -  \Xc ) s_{\vartheta_S }^2  \Big) Z_m^\prime   \Big\} \cdot \Psi_{ \rep{3} } \,, \label{eq:331_ZprimeCov_fund}\\[2mm]
%
d_m \Psi_{ \repb{3} }&\equiv& \Big\{ \partial_m \mathbb{I}_3  - i g_{Z^\prime } {\rm diag} \Big( (  - \frac{1}{6} +  ( \frac{1}{6} - \bar  \Xc ) s_{\vartheta_S }^2  ) \mathbb{I}_{2  }  \,,  \frac{1}{3} - (\frac{1}{3} + \bar \Xc ) s_{\vartheta_S }^2  \Big) Z_{{\rm ANO}\, m}^\prime   \Big\} \cdot \Psi_{ \repb{3} } \,,\label{eq:331_ZprimeCov_antifund}
\eeqn
%\eeqs
%
%
with $m=(1\,,2)$ representing the two-dimensional Cartesian coordinates.

\para
The $Z^\prime$ string tension is given by
\beqn\label{eq:331_tension}
\mu_{\rm ANO}&=& \int  d^2 x \, \Big\{  \frac{1}{4} ( W_{ {\rm ANO}\, m n}^I )^2 +   \frac{1}{4} ( X_{ {\rm ANO}\, mn } )^2  + | d_m \Phi_{ {\rm ANO}\, \repb{3}\,, \omega } |^2 + V(  \Phi_{{\rm ANO}\, \repb{3}\,, \omega })   \Big\} \,,
\eeqn
with the covariant derivatives in the $Z^\prime$ string background defined in Eq.~\eqref{eq:331_ZprimeCov_antifund}, and the 331 Higgs potential $V(  \Phi_{ \repb{3}\,, \omega})$ given in Eq.~\eqref{eq:331_potential}.
Explicitly, we have
\beqs
\beqn
&& \frac{1}{4} \( W_{{\rm ANO}\, mn }^{I =8}  \)^2 + \frac{1}{4} \( X_{ {\rm ANO}\, mn }  \)^2 = \hf \( \partial_{ [1 } Z_{ {\rm ANO}\, 2] }^\prime  \)^2 =   \hf \( - \frac{1 }{  r } \frac{d \bar \zeta (r ) }{d r} \)^2 \,, \\[2mm]
%
%&& d_m  \Phi_{ {\rm ANO}\, \repb{3}\,, \omega } =  \Big\{ \partial_m \mathbb{I}_3  - i g_{Z^\prime } {\rm diag} \Big( (  - \frac{1}{6} +   \frac{1}{2}  s_{\vartheta_S }^2  ) \mathbb{I}_{2  }  \,,  +\frac{1}{3}   \Big) Z_{{\rm ANO} \,,m}^\prime   \Big\}  \cdot   \left( \ba{c}  0 \\    0  \\   \phi_{ {\rm ANO} \,, \omega }  \ea  \right) \non
%&=& \frac{ 1  }{ \sqrt{2}} V_\omega \( \partial_m ( \bar f_\omega (r) e^{i \varphi } ) - \frac{ i }{ 3} g_{Z^\prime } Z_{{\rm ANO}\,, m }^\prime (\bar f_\omega  (r) e^{i \varphi } )  \)  \left( \ba{c}  0 \\    0  \\   1 \ea  \right)  \non
%
&& | d_m  \Phi_{ {\rm ANO}\, \repb{3}\,, \omega }  |^2 = \hf V_{\omega }^2 \Big[ ( \frac{d \bar f_\omega (r) }{ d r} )^2 + \( 1  + \frac{ g_{Z^\prime }   }{3 }  \bar \zeta (r )  \)^2 \( \frac{\bar f_\omega (r ) }{r }  \)^2 \Big] \,,\\[2mm]
&& V( \Phi_{ {\rm ANO}\, \repb{3}\,, \omega }  ) =   \frac{  1}{8 } \lambda_1V_1^4 \Big( \bar f_1^2 (r) - 1  \Big)^2  + \frac{  1 }{8 } \lambda_2 V_2^4 \Big( \bar f_2^2 (r)  - 1  \Big)^2 \non
&+& \f{1}{8} \lambda_3 \Big( V_1^2 \bar f_1^2 (r) + V_2^2 \bar f_2^2 (r) - v_{331 }^2  \Big)^2   + \frac{ 1 }{4 } \lambda_{5} V_1^2 V_2^2 \Big( \bar f_1 (r) \bar f_2 (r) - 1  \Big)^2 \,.
\eeqn
\eeqs
With the dimensionless coordinates of
\beqn\label{eq:xi_coord}
&& \xi \equiv \f{ m_{Z^\prime } }{ \sqrt{2 } } r= \f{ g_{Z^\prime } }{ 3 \sqrt{2} } v_{331} r \,,
\eeqn
the $Z^\prime$ string tension is expressed as
\beqn\label{eq:331_tension_dimensionless}
%\mu_{\rm ANO}&=& 2 \pi \int   r d r \, \Big\{  \frac{1}{ 2 r^2 } \( \frac{d \bar \zeta (r ) }{dr } \)^2 + \hf \sum_{\omega=1\,,2} V_{\omega }^2 \Big( ( \frac{d \bar f_\omega (r) }{ d r} )^2 + ( 1  + \frac{ g_{Z^\prime }  }{3 }  \bar \zeta (r ) )^2 (\frac{\bar f_\omega (r ) }{r } )^2 \Big)  \non
%&+&   \frac{  1}{8 } \lambda_1 v_{331 }^4 c_{\tilde \beta }^4  \Big( \bar f_1^2 (r) - 1  \Big)^2  + \frac{  1 }{8 } \lambda_2  v_{331 }^4 s_{\tilde \beta }^4  \Big( \bar f_2^2 (r)  - 1  \Big)^2 \non
%&+& \f{1}{8} \lambda_3 v_{331 }^4 \Big( c_{ \tilde \beta }^2 \bar f_1^2 (r) + s_{ \tilde \beta }^2 \bar f_2^2 (r) - 1   \Big)^2   + \frac{ 1 }{4 } \lambda_{5} v_{331 }^4 s_{\tilde \beta }^2 c_{\tilde \beta }^2   \Big( \bar f_1 (r) \bar f_2 (r) - 1  \Big)^2 \Big\}  \non
%
\mu_{\rm ANO}&=& 2\pi v_{331}^2  \int \xi d \xi \,  \rho(\xi )   \,,  \quad \rho(\xi )=  \rho_{ \bar \zeta }( \xi) + \rho_{ \bar f_\omega }( \xi ) + \rho_V( \xi ) \,, \non
\rho_{ \bar \zeta}(\xi) &=&    \frac{  g_{Z^\prime}^2 }{ 36  } \( \f{1}{ \xi } \frac{d \bar \zeta ( \xi ) }{d \xi } \)^2  \,, \non
\rho_{ \bar f_\omega }(\xi)&=&  \hf \sum_{\omega=1\,,2} \f{ V_{\omega }^2 }{ v_{331}^2 } \Big[  \(  \frac{d \bar f_\omega ( \xi ) }{ d  \xi } \)^2 + \( 1  + \frac{ g_{Z^\prime }  }{3 }  \bar \zeta ( \xi ) \)^2 \( \frac{\bar f_\omega ( \xi ) }{ \xi }  \)^2 \Big]  \,, \non
\rho_V ( \xi ) &=&  \frac{  1}{4 } \beta_1 c_{\tilde \beta }^4  \Big( \bar f_1^2 ( \xi ) - 1  \Big)^2  + \frac{  1 }{4 } \beta_2   s_{\tilde \beta }^4  \Big( \bar f_2^2 ( \xi )  - 1  \Big)^2 \non
&&+ \f{1}{4 } \beta_3   \( c_{ \tilde \beta }^2 \bar f_1^2 ( \xi ) + s_{ \tilde \beta }^2 \bar f_2^2 ( \xi ) - 1   \)^2   +  \frac{ 1 }{2 } \beta_{5}   s_{\tilde \beta }^2 c_{\tilde \beta }^2   \( \bar f_1 ( \xi ) \bar f_2 ( \xi ) - 1  \)^2   \,,
\eeqn
where the $( \rho_{ \bar \zeta }( \xi) \,, \rho_{ \bar f_\omega }( \xi ) \,, \rho_V( \xi ))$ represent the dimensionless energy densities from the gauge field kinematic terms, the Higgs field kinematic terms, and the Higgs potential.
In the energy density from the Higgs potential $\rho_V ( \xi )$, we parametrized the ratios between the Higgs self-couplings and the gauge coupling as
\beqn\label{eq:331_betai}
&& \beta_i \equiv \frac{9 \lambda_i }{  g_{Z^\prime}^2 } \,.
\eeqn

\subsection{The $Z^\prime$ string solutions}
\label{section:string_sol}

 %%%%%%%%%%%%%%%%%%%%%%%%%%%%%%%%%%%%%%%%%%%%%%%%%%%%%%%
\begin{figure}[htb]
\centering
\includegraphics[height=4.8cm]{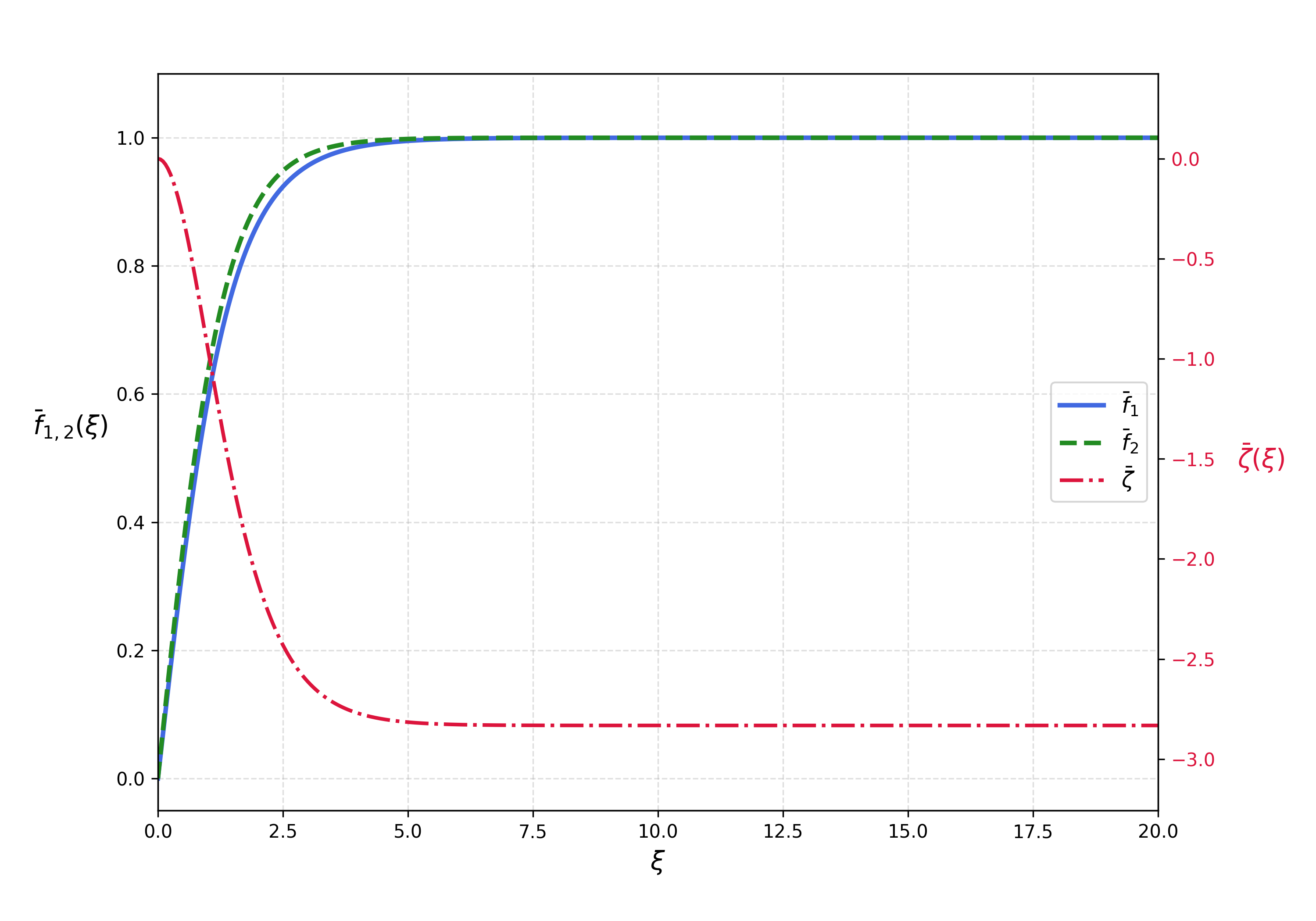}
\includegraphics[height=4.8cm]{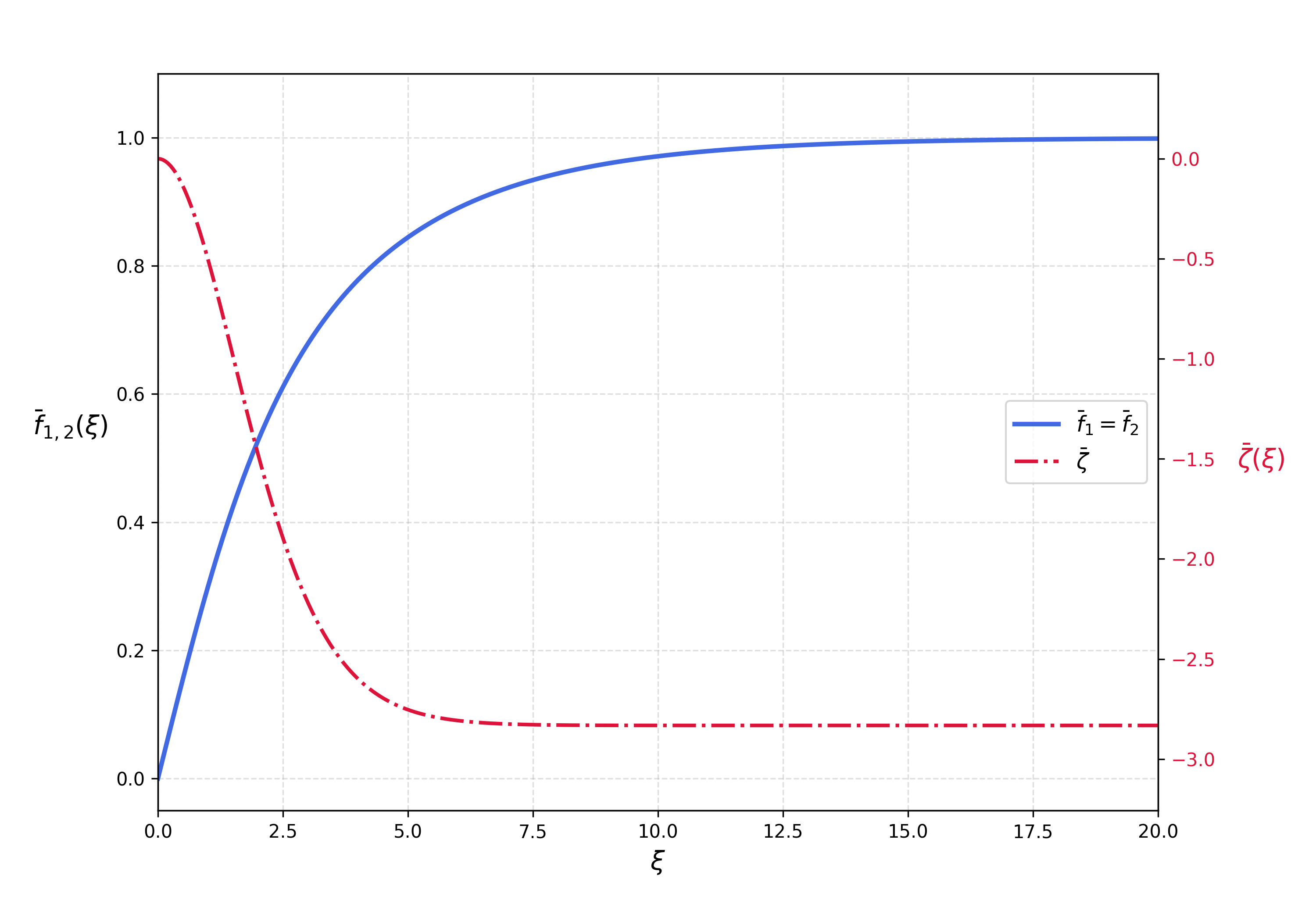}
\includegraphics[height=4.8cm]{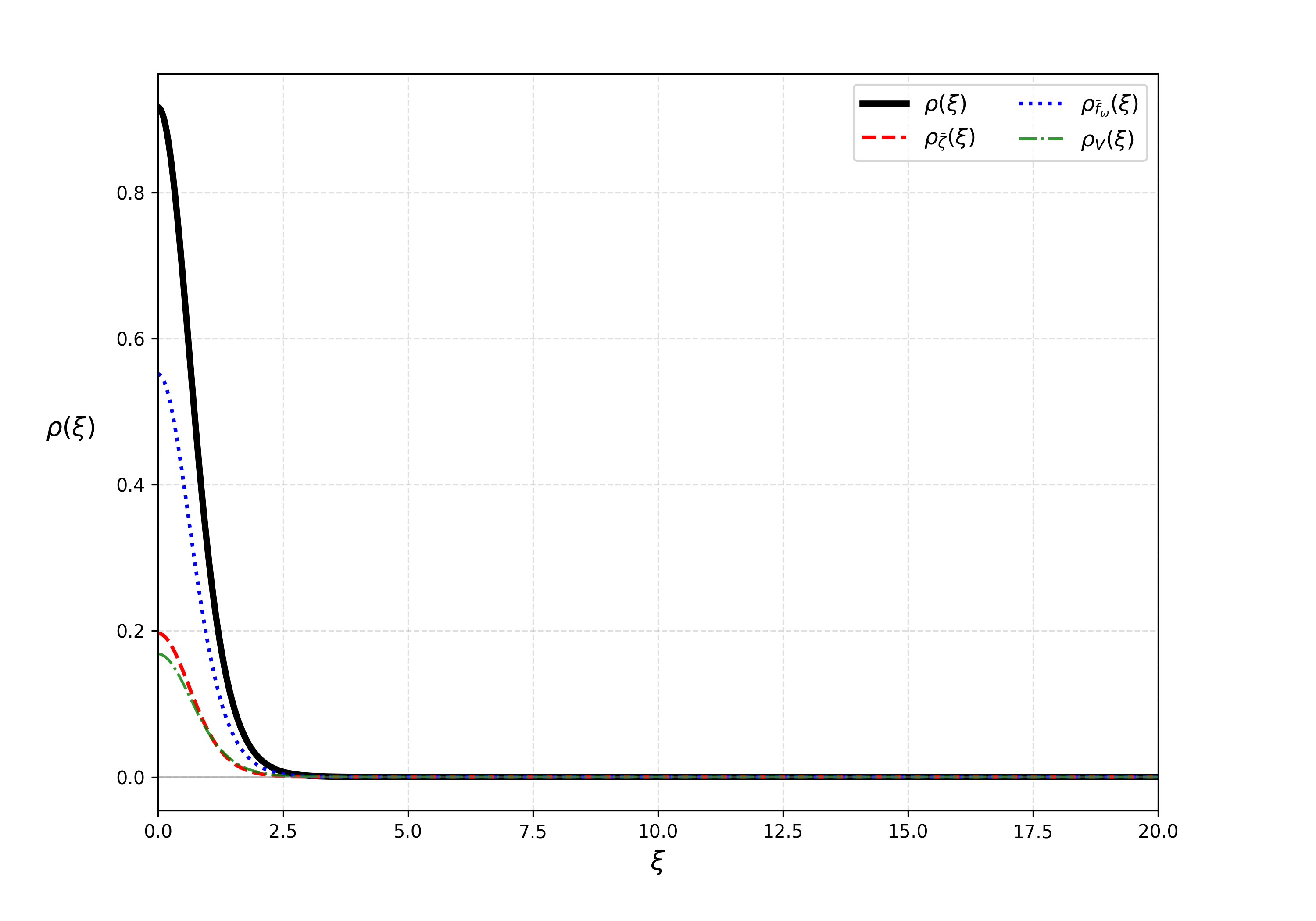}
\includegraphics[height=4.8cm]{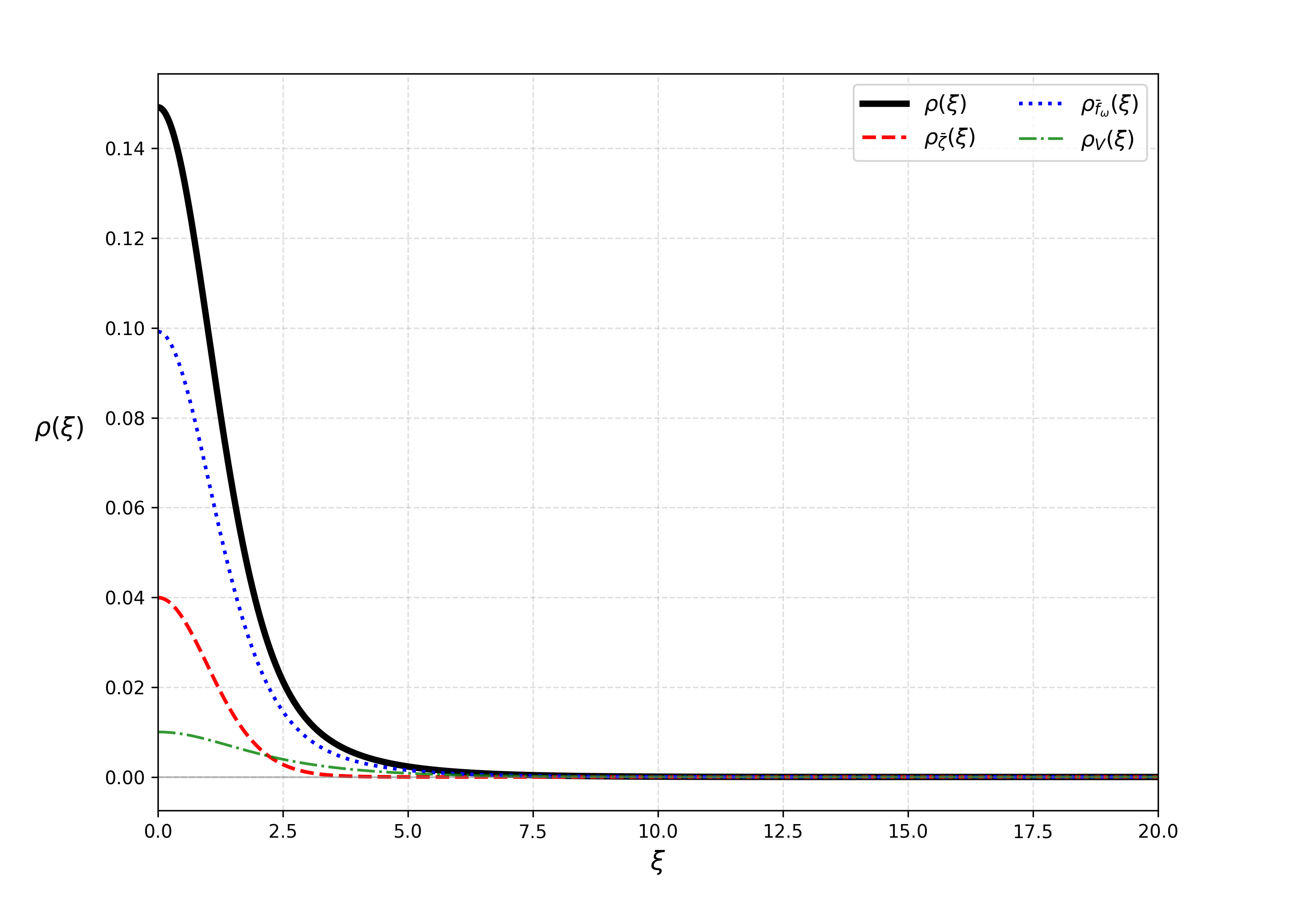}
\caption{The 331 $Z^\prime$ string profile functions of $(\bar f_1(\xi)\,, \bar f_2 (\xi)\,, \bar \zeta(\xi) )$ (upper panels) and the dimensionless string energy densities (lower pannels).
The parameters for the left panels are: $\lambda_1=\lambda_2=0.15$, $\lambda_3=-0.05$, $\lambda_5=0.1$, $t_{\tilde \beta}=2$, and $g_{Z^\prime}=1.0592$.
The parameters for the right panels are: $\lambda_1=\lambda_2= \lambda_5 = 0.15$, $\lambda_3=-0.145$, $t_{\tilde \beta}=1$, and $g_{Z^\prime}=1.0592$.
}
\label{fig:331_profiles} 
\end{figure}
%%%%%%%%%%%%%%%%%%%%%%%%%%%%%%%%%%%%%%%%%%%%%%%%%%%%%%%

\para
The classical Euler-Lagrange field equations are
\beqs\label{eqs:331string_EOM_dimensionless}
\beqn
&&  \f{ d^2 \bar f_1 (\xi )}{d \xi^2} + \f{1}{ \xi } \f{ d \bar f_1 (\xi )}{d \xi }   - \( 1 +\f{ g_{Z^\prime } }{ 3} \bar \zeta ( \xi ) \)^2 \f{ \bar f_1 (\xi ) }{ \xi^2 }  + \beta_1 c_{\tilde \beta }^2  \( 1 - \bar f_1^2 ( \xi )  \) \bar f_1 ( \xi )   \non 
&& + \beta_3 \( 1-  c_{\tilde \beta}^2 \bar f_1^2 (\xi ) - s_{\tilde \beta}^2 \bar f_2^2 (\xi )  \)  \bar f_1 (\xi ) + \beta_5  s_{\tilde \beta}^2   \(  1- \bar f_1 ( \xi ) \bar f_2 ( \xi )  \) \bar f_2 ( \xi ) =0  \,, \label{eq:331string_EOM_f1}\\[2mm]
%%%%%%%%%%%%%%%%%%%%%%%%%%%%%%%%%%%%%%%%%%%%%
&&  \f{ d^2 \bar f_2 ( \xi )}{d \xi^2}  +  \f{1}{ \xi } \f{ d \bar f_2 ( \xi )}{d \xi }   -  \( 1 +\f{ g_{Z^\prime } }{ 3} \bar \zeta ( \xi ) \)^2 \f{ \bar f_2 (\xi ) }{ \xi^2 }  + \beta_2 s_{\tilde \beta }^2  \( 1- \bar f_2^2 ( \xi )  \) \bar f_2 (\xi ) \non
&&  + \beta_3  \( 1- c_{\tilde \beta}^2 \bar f_1^2 (\xi ) -  s_{\tilde \beta}^2 \bar f_2^2 ( \xi )  \)  \bar f_2 ( \xi ) + \beta_5 c_{\tilde \beta}^2 \( 1 -  \bar f_1 (  \xi ) \bar f_2 ( \xi )  \) \bar f_1 ( \xi )  =0 \,, \label{eq:331string_EOM_f2}\\[2mm]
%%%%%%%%%%%%%%%%%%%%%%%%%%%%%%%%%%%%%%%%%%%%%
&&  -  \f{ d^2 \bar \zeta(\xi) }{d \xi^2 } + \f{1}{ \xi } \f{d \bar \zeta( \xi )}{d \xi }  + \f{6 }{ g_{Z^\prime } } \(1 + \f{ g_{Z^\prime} }{3 }  \bar \zeta( \xi ) \) \( c_{\tilde \beta}^2 \bar f_1^2 ( \xi ) +  s_{\tilde \beta }^2 \bar f_2^2 ( \xi )  \) =0  \,.\label{eq:331string_EOM_zeta}
\eeqn
\eeqs
The Dirichlet boundary conditions to Eqs.~\eqref{eqs:331string_EOM_dimensionless} are
\beqn\label{eqs:331string_boundary}
&& \bar f_1(\xi =0) = \bar f_2( \xi= 0)= \bar \zeta(\xi= 0) = 0 \,, \non
&&  \bar f_1( \xi= \infty ) =  \bar f_2 ( \xi= \infty ) = 1 \,, ~~ \bar \zeta( \xi= \infty)= - \frac{3}{ g_{Z^\prime }} \,.
\eeqn
At $\xi\to \infty$, the asymptotic behaviors for the Higgs profiles read
\beqn\label{eq:331_Higgs_asymptotic}
&& \bar f_1 ( \xi) \sim 1 - \f{1}{ \sqrt{\xi} } \( C_1 \exp ( - \sqrt{ \beta_1 + 2 \beta_3 + \beta_5 } \xi )  + C_2  \exp ( - \sqrt{\beta_1 } \xi ) \) \,, \non
&& \bar f_2 ( \xi) \sim 1 - \f{1}{ \sqrt{\xi} } \( C_1 \exp ( - \sqrt{ \beta_1 + 2 \beta_3 + \beta_5 } \xi )  - C_2  \exp ( - \sqrt{\beta_1 } \xi ) \) \,,
\eeqn
with $\tilde \beta =\f{\pi}{4}$, $\beta_1=\beta_2$, and $C_{1\,,2}$ being some constants.

\para
In Fig.~\ref{fig:331_profiles}, we display two $Z^\prime$ string profile functions as well as the distributions of the dimensionless energy densities with two different parameter inputs based on Eqs.~\eqref{eqs:331string_EOM_dimensionless} and \eqref{eqs:331string_boundary}.
In all these plots, we vary the dimensionless distances within the range of $\xi\in(0\,,20)$.
Among them, the contributions to the string tension from the Higgs potential are always dominant over the kinematic terms of the gauge and Higgs fields in both plots.
We take different inputs of $(\lambda_3\,,t_{\tilde \beta})=( -0.05 \,, 2)$ in the left panels and $(\lambda_3\,,t_{\tilde \beta})=( -0.145 \,, 1)$ in the right panels, while keep other inputs to be the same for comparison.
The negative values of $\lambda_3$ are taken in order to alleviate the instability originated from the Higgs potential.
Meanwhile, $\lambda_3$ can not take arbitrary negative value with the boundary conditions in Eq.~\eqref{eqs:331string_boundary}.
To see this, let us take the simplified case of $t_{\tilde \beta}=1$ and $\lambda_1=\lambda_2=\lambda_5=0.15$ in the right panels of Fig.~\ref{fig:331_profiles}~\footnote{The values of $(\lambda_1\,, \lambda_2)>0$ or equivalently $(\beta_1\,, \beta_2)>0$ should be satisfied in order to bound the Higgs potential in Eq.~\eqref{eq:331_potential} from below.}, where the evolutions of two Higgs profile functions of $( \bar f_1(\xi) \,, \bar f_2( \xi))$ in Eqs.~\eqref{eq:331string_EOM_f1} and \eqref{eq:331string_EOM_f2} are identical.
The Higgs potential contributions in Eq.~\eqref{eq:331string_EOM_f1} are thus reduced to $\sim ( \hf \beta_1 + \hf \beta_5 + \beta_3) \(1 - \bar f^2 ( \xi) \) \bar f(\xi)$, with $\bar f_1 (\xi )= \bar f_2(\xi) \equiv \bar f(\xi )$.
Together with the asymptotic forms of the Higgs fields in Eq.~\eqref{eq:331_Higgs_asymptotic}, one expects the relation of $  \beta_3> - \hf ( \beta_1 + \beta_5 )$ from Eq.~\eqref{eq:331string_EOM_f1} or $\beta_3> - \hf ( \beta_2 + \beta_5 )$ from Eq.~\eqref{eq:331string_EOM_f2} to hold for this simplified case.
Otherwise, one would expect a wave-like behavior of the Higgs profile functions at $\xi=\infty$, which are unwanted for the $Z^\prime$ string.
For the detailed string stability analysis in Sec.~\ref{section:331_details}, we will demonstrate one such example with some positive correction to the lower bound of the $\beta_3$ in order to avoid the numerical uncertainties.

%###################################################################
\vspace*{3mm}
\section{The classical stability of the $Z^\prime$ string}
\label{section:stability_331}
%###################################################################

\subsection{The perturbations to the $Z^\prime$ string}
\label{section:331_perturbation}

\para
The classical stability analyses for the $Z$ string in the SM and beyond have been carried out in Refs.~\cite{James:1992zp,James:1992wb,Earnshaw:1993yu,Goodband:1995rt,Goodband:1995he,Kanda:2022xrz,Eto:2024xvc}.
Similar to the approaches in Refs.~\cite{Goodband:1995rt,Goodband:1995he,Eto:2024xvc}, we impose the small perturbations~\footnote{We denote the complex conjugates of the perturbed scalar fields as $( \delta \pi_{W^\prime\,, \omega }^- )^* \equiv  \delta \pi_{W^\prime\,, \omega }^+$ and $( \delta \pi_{N  \,, \omega }^0 )^* \equiv \delta \pi_{ \bar N  \,, \omega }^0$.} to the string background in Eq.~\eqref{eq:331_string_form} by following Eqs.~\eqref{eq:331_Higgs} and \eqref{eq:331_connection_antifund}
\beqn\label{eq:331_string_perturb}
&& \Phi_{ \repb{ 3 } \,, \omega }  =  \left( 
\ba{c}  
 \delta \pi_{W^\prime\,, \omega }^-    \\ 
 \delta \pi_{\bar N  \,, \omega }^0 \\ 
  \phi_{{\rm ANO} \,, \omega }  \\  \ea  \right) \,,\quad \omega=(1\,, 2) \,, \non
%
%D_m  \Phi_{ \repb{ 3 } \,, \omega } &=& \Big( \partial_m \mathbb{I}_3  +i  \frac{g_{Z^\prime } c_{ \vartheta_S } }{ \sqrt{6 } } \left( 
%\ba{ccc}  
% 0  & 0  & \delta W_m^{\prime -}  \\ 
%0  &  0 & \delta \bar N_m  \\ 
%\delta W_m^{\prime\, +}   &  \delta  N_m  &  0  \\  \ea  \right)  \non
%&& - i g_{Z^\prime }  {\rm diag}  (  ( - \frac{1}{6} +  \frac{1}{2 }   s_{\vartheta_S }^2  ) \mathbb{I}_{2 }  \,,  + \frac{1}{3}  ) Z_{{\rm ANO} \,, m}^\prime  \Big) \cdot  \Phi_{ \repb{ 3 } \,, \omega }  \non
%
D_m  \Phi_{ \repb{ 3 } \,, \omega }&=& \left( 
\ba{c}   
 \partial_m \delta \pi_{W^\prime\,, \omega }^-  +  \frac{ i g_{ Z^\prime } c_{\vartheta_S } }{ \sqrt{6} }  \phi_{\rm ANO}   \delta W_m^{\prime\, - }   - i g_{Z^\prime} ( - \frac{1 }{ 6} + \hf s_{ \vartheta_S }^2 ) Z_{ {\rm ANO}\,, m }^\prime \delta \pi_{ W^\prime\,, \omega }^- \\[2mm] 
 \partial_m \delta \pi_{\bar N  \,, \omega }^0 +  \frac{ i g_{ Z^\prime } c_{\vartheta_S } }{ \sqrt{6} }  \phi_{\rm ANO}   \delta \bar N_m  - i g_{Z^\prime} ( - \frac{1 }{ 6} + \hf s_{ \vartheta_S }^2 ) Z_{ {\rm ANO}\,, m }^\prime  \delta \pi_{ \bar N  \,, \omega }^0  \\[2mm] 
  \partial_m  \phi_{\rm ANO}   -  \frac{ i g_{Z^\prime } }{3 } Z_{ {\rm ANO}\,, m }^\prime  \phi_{\rm ANO}  +   \frac{ i g_{ Z^\prime } c_{\vartheta_S } }{ \sqrt{6} } ( \delta W_m^{\prime\, + } \delta \pi_{ W^\prime\,, \omega }^-  + \delta N_m \delta \pi_{ \bar N \,,  \omega }^0 )  \\
    ~~~   \\   \ea \right)\,.
\eeqn
%
%
\begin{comment}
The following relations are useful
%
%
\beqn
%
&& W_{m n }^{ \prime\, \pm } = \frac{1}{ \sqrt{2} } ( W_{mn }^4  \mp i W_{mn }^5  )  =  \partial_{ [ m } \delta W_{ n ]}^{\prime\, \pm }  \pm  \frac{ i g_{Z^\prime } }{  2  }  c_{ \vartheta_S }^2 \delta W_{[ m }^{\prime\, \pm } Z_{ {\rm ANO} \,, n ]  }^\prime  \,, \non
%
&& N_{mn}  = \frac{1}{ \sqrt{2} } ( W_{mn }^6  - i W_{mn }^7  ) = \partial_{ [ m } \delta N_{ n ]}  +  \frac{ i g_{Z^\prime } }{ 2  }  c_{ \vartheta_S }^2 \delta N_{[ m } Z_{ {\rm ANO} \,, n ]  }^\prime  \,, \non
%
&& \bar N_{m n } = \frac{1}{ \sqrt{2} } ( W_{mn }^6  +  i W_{mn }^7  ) =  \partial_{ [ m } \delta \bar N_{ n ]}  -  \frac{ i  g_{Z^\prime }}{  2  }  c_{ \vartheta_S }^2 \delta \bar N_{[ m } Z_{ {\rm ANO} \,, n ]  }^\prime   \,, \non
%&& W_{ m n }^\pm = \frac{1}{ \sqrt{2} } (  W_{mn }^1 \mp i W_{ mn}^2  ) =\partial_{[m} \delta W_{ n]}^\pm  \pm i g c_W \delta W_{ [m}^\pm Z_{ {\rm ANO}\,, n] }\,,\non
%
&& W_{mn }^8 = c_{ \vartheta_S } \partial_{ [ m }  Z_{ {\rm ANO}\,, n ]}^\prime  -  \frac{ i g_{Z^\prime }}{2 }  c_{ \vartheta_S } (    \delta W^{\prime\, +}_{ [ m } \delta W^{\prime\, -}_{ n ] }   +    \delta  N_{ [ m } \delta \bar N_{n ] }    )   \,.
%
\eeqn
\end{comment}
%
%
Accordingly, we expand the string tension to the quadratic terms as follows
\beqn
\mu&=& \mu_{\rm ANO} + \delta \mu_{W^\prime } + \delta \mu_\pi + \delta \mu_c  \,.
\eeqn
Each perturbed string tension must include two sets of identical contributions from the perturbed $( \delta W_m^{\prime\, \pm } \,,  \delta \pi_{W^\prime\,, \omega }^\pm )$ and $(\delta N_m \,, \delta \bar N_m \,, \delta \pi _{ N\,, \omega}^0 \,, \delta \pi_{ \bar N \,, \omega}^0)$, respectively.
This is due to the identical ${\rm U}(1)_Y$ hypercharges as shown in Eq.~\eqref{eq:331Y_offGB}.
It is therefore sufficient to consider the perturbations involving the fields of $( \delta W_m^{\prime\, \pm } \,,  \delta \pi_{W^\prime\,, \omega }^\pm )$ below.

\para
Now, we display the expressions for the perturbed string tension in terms of the dimensionless coordinates given in Eq.~\eqref{eq:xi_coord} explicitly.
The $\delta \mu_{W^\prime}$ term comes from the perturbations to the kinematic terms of the gauge fields 
\beqn\label{eq:gauge_perturb}
 \delta \mu_{W^\prime }&=& \int d^2 \xi \, \Big\{   \[ \partial_{[ \xi_1 } \delta  W_{ \xi_2 ]}^{\prime\, + } -  \f{ i g_{Z^\prime } c_{\vartheta_S }^2 }{2 } \f{ \bar \zeta ( \xi ) }{ \xi }  \(  \delta  W_{ \xi_1 }^{\prime\, + } c_\varphi + \delta  W_{ \xi_2  }^{\prime\, + } s_\varphi  \) \] \non
&& \cdot \[ \partial_{[ \xi_1 } \delta W_{ \xi_2 ]}^{ \prime\, - } + \f{  i g_{Z^\prime } c_{\vartheta_S }^2 }{2 } \f{ \bar \zeta ( \xi ) }{  \xi  }  \(  \delta  W_{ \xi_1 }^{\prime\, - } c_\varphi  + \delta  W_{ \xi_2 }^{\prime\, - } s_\varphi \) \] \non
%
%&& + \left( \partial_{[ \xi_1 } \delta  N_{ 2 ]} -  \f{ i g_{Z^\prime } c_{\vartheta_S }^2 }{2 } \f{ \bar \zeta ( \xi ) }{ \xi }  \(  \delta  N_{ 1 } c_\varphi + \delta  N_{2  } s_\varphi  \) \right) \non
%&& \cdot \left( \partial_{[ \xi_1 } \delta \bar N_{ 2 ]} + \f{ i g_{Z^\prime } c_{\vartheta_S }^2 }{2 }  \f{ \bar \zeta ( \xi ) }{  \xi  }  \(  \delta \bar N_{ 1 } c_\varphi  + \delta \bar N_{ 2 } s_\varphi \) \right) \non
%
&&   + \f{ i g_{Z^\prime } c_{\vartheta_S }^2 }{2 }  \f{1}{ \xi } \f{ d \bar \zeta( \xi ) }{d \xi }  \delta W_{[ \xi_1 }^{\prime\, + } \delta  W_{ \xi_2 ]}^{ \prime \, -}   +  \f{ 3 c_{\vartheta_S }^2 }{2 }  \( \sum_\omega \f{ V_\omega^2 }{  v_{331}^2 }    \bar f_\omega^2 (\xi ) \)\delta  W_{\xi_m}^{ \prime\, + }  \delta W_{\xi_m}^{ \prime\, -}    \Big\} \,.
%&=&  2 \pi \int r d r\,  \Big\{    \[ ( \vec \nabla \times \delta \vec W^{\prime\, + } )_z - \frac{ i g_{Z^\prime }}{  2  }  c_{ \vartheta_S }^2 \frac{\bar \zeta (r) }{r } \delta W_r^{\prime\, + }  \] \cdot \[  ( \vec \nabla \times \delta \vec W^{\prime\, - } )_z  + \frac{ i g_{Z^\prime } }{ 2 }  c_{ \vartheta_S }^2 \frac{\bar \zeta (r) }{r } \delta W_r^{\prime\, - }   \] \non
%&& +  \[ ( \vec \nabla \times \delta \vec N )_z -  \frac{ i g_{Z^\prime } }{ 2 }  c_{ \vartheta_S }^2 \frac{\bar \zeta (r) }{r } \delta N_r  \] \cdot \[  ( \vec \nabla \times \delta \vec{ \bar N } )_z  + \frac{ i  g_{Z^\prime } }{ 2 } c_{ \vartheta_S }^2 \frac{\bar \zeta (r) }{r } \delta \bar N_r  \]  \non
%&& + \frac{i  g_{Z^\prime } }{ 2}  c_{ \vartheta_S }^2 \frac{1}{r} \frac{ d \bar \zeta (r) }{ dr }  \cdot ( \delta W_{[ r}^{\prime\, + } \delta W_{\varphi ] }^{\prime\, - }  + \delta  N_{ [ r } \delta \bar N_{\varphi ] }  ) \Big\}  \,,
\eeqn
%
%
%where the $z$-component of the curl operator reads
%
%
%\beqn
%&& ( \vec \nabla \times \delta \vec \Ac )_z = \f{ \partial \delta \Ac_\varphi }{ \partial r } + \f{1}{r } (  \delta \Ac_\varphi - \f{ \partial \delta \Ac_r }{ \partial \varphi } ) \,.
%\eeqn
%
%
The $\delta \mu_\pi$ term comes from the covariant derivative terms and the Higgs potential only containing the perturbed scalar components
\beqn\label{eq:scalar_perturb}
 \delta \mu_\pi &=&  \sum_\omega \int  d^2 \xi  \, \Big\{   \Big| \Big( \partial_{\xi_1} - i g_{Z^\prime } \( - \frac{1}{6} + \hf s_{\vartheta_S }^2 \)  \( \f{\bar \zeta ( \xi)}{\xi } s_\varphi   \)  \Big) \delta \pi_{W^\prime\,, \omega }^-  \Big|^2 \non
 && +  \Big|  \Big( \partial_{\xi_2 } - i g_{Z^\prime } \( - \frac{1}{6} + \hf s_{\vartheta_S }^2 \)  \( - \f{\bar \zeta ( \xi)}{\xi } c_\varphi   \)   \Big) \delta \pi_{W^\prime\,, \omega }^-  \Big|^2 + V( \delta \pi_{W^\prime\,, \omega }^\pm  )    \Big\} \,,
\eeqn
where
\beqn\label{eq:331potential_perturb}
&&\sum_\omega V(  \delta \pi_{W^\prime  \,, \omega }^\pm    )  =  \beta_1  c_{\tilde \beta}^2  \( \bar f_1^2 ( \xi )  - 1 \) \delta \pi_{W^\prime  \,, 1 }^+ \delta \pi_{W^\prime  \,, 1 }^- +  \beta_2  s_{\tilde \beta}^2  \( \bar f_2^2 ( \xi )  - 1 \) \delta \pi_{W^\prime  \,, 2 }^+ \delta \pi_{W^\prime  \,, 2 }^-    \non
&&   +  \beta_3 \(  c_{\tilde \beta}^2 ( \bar f_1^2( \xi )- 1 ) +   s_{\tilde \beta}^2  ( \bar f_2^2( \xi )- 1 ) \) \( \delta \pi_{W^\prime  \,, 1 }^+ \delta \pi_{W^\prime  \,, 1 }^-  + \delta \pi_{W^\prime  \,, 2 }^+ \delta \pi_{W^\prime  \,, 2 }^-   \)   \non
&&  +  \beta_4  \(  s_{\tilde \beta}^2 \bar f_2^2 ( \xi ) \delta \pi_{W^\prime  \,, 1 }^+ \delta \pi_{W^\prime  \,, 1 }^-   + c_{\tilde \beta}^2 \bar f_1^2 ( \xi ) \delta \pi_{W^\prime  \,, 2 }^+ \delta \pi_{W^\prime  \,, 2 }^-  -   s_{\tilde \beta} c_{\tilde \beta} \bar f_1( \xi ) \bar f_2 ( \xi )  ( \delta \pi_{W^\prime  \,, 1 }^-  \delta \pi_{W^\prime  \,, 2 }^+ + \delta \pi_{W^\prime  \,, 1 }^+  \delta \pi_{W^\prime  \,, 2 }^- )  \)   \non
&&  + \beta_5  s_{\tilde \beta} c_{\tilde \beta} \( \bar f_1( \xi ) \bar f_2 ( \xi ) -  1  \) \( \delta \pi_{W^\prime  \,, 1 }^-  \delta \pi_{W^\prime  \,, 2 }^+  + \delta \pi_{W^\prime  \,, 1 }^+ \delta \pi_{W^\prime  \,, 2 }^-  \) \,.
\eeqn
The $\delta \mu_c$ term comes from the covariant derivative terms involving the both the scalar perturbations and the gauge field perturbations
\beqn\label{eq:scalar_gauge_perturb}
\delta \mu_c &=&  i  c_{\vartheta_S } \sqrt{ \f{ 3  }{   2 }   } \sum_\omega  \frac{ V_\omega }{  v_{331} }    \int d^2 \xi  \Big\{  \bar f_\omega ( \xi )  \( e^{i\varphi } ( \partial_{\xi_m} \delta \pi_{W^\prime \,, \omega }^+   )  \delta  W_{\xi_m}^{\prime\, - }  -  e^{ - i\varphi }  ( \partial_{\xi_m } \delta \pi_{W^\prime \,, \omega }^-   )  \delta  W_{\xi_m}^{\prime\, + } \)   \non
%
%&& + \bar f_\omega ( \xi )  \( e^{i\varphi }  \partial_{\xi_m} \delta \pi_{N \,, \omega }^0   \cdot  \delta \bar N_m  -  e^{ - i\varphi }   \partial_{\xi_m } \delta \pi_{\bar N \,, \omega }^0  \cdot  \delta N_m \)  \non
%
&& +  \partial_{\xi_m } ( \bar f_\omega (\xi ) e^{ - i\varphi } )   \cdot \delta  W_{\xi_m}^{\prime\, +} \delta \pi_{W^\prime \,, \omega }^-  -   \partial_{\xi_m } ( \bar f_\omega ( \xi ) e^{i\varphi } )  \cdot \delta  W_{\xi_m}^{ \prime\, - } \delta \pi_{W^\prime \,, \omega }^+   \non
%&& +  \partial_{\xi_m } ( \bar f_\omega (\xi ) e^{ - i\varphi } )   \cdot \delta  N_m \delta \pi_{\bar N \,, \omega }^0  -   \partial_{\xi_m } ( \bar f_\omega ( \xi ) e^{i\varphi } )  \cdot \delta \bar N_m \delta \pi_{N \,, \omega }^0   \non
%
&& +  i g_{Z^\prime } (  \frac{1}{6 } + \hf s_{ \vartheta_S }^2  )   \bar f_\omega ( \xi ) \f{ \bar \zeta (\xi ) }{\xi }   \Big[  s_\varphi   \( e^{ - i\varphi }    \delta  W_{\xi_1}^{\prime\, + } \delta \pi_{W^\prime \,, \omega }^-  +  e^{i\varphi }  \delta  W_{\xi_1}^{ \prime \, - } \delta \pi_{ W^\prime \,, \omega }^+   \) \non
&&   -  c_\varphi  \(e^{ - i\varphi }    \delta  W_{\xi_2}^{\prime\, + } \delta \pi_{W^\prime \,, \omega }^-  +  e^{i\varphi }  \delta  W_{\xi_2}^{ \prime \, - } \delta \pi_{ W^\prime \,, \omega }^+   \)  \Big]  \Big\} \,.
\eeqn
Similar to the $Z$ string perturbation in the SM~\cite{Goodband:1995rt,Goodband:1995he}, the first line in Eq.~\eqref{eq:scalar_gauge_perturb} contains linear derivative terms and can be removed by the gauge fixing terms below.

\subsection{The gauge fixing terms}
\label{section:331_gf}

\para
The gauge fixing terms for the 331 model expressed in the dimensionless Cartesian coordinates are~\footnote{The terms from the Higgs fields in the gauge fixing terms differ from the convention in Ref.~\cite{Goodband:1995he} by a minus sign, since our Higgs fields are in the anti-fundamental representation of ${\rm SU}(N)$.}
\beqn\label{eq:331_fix}
\Lc_{\rm fix}&=&  F(   \delta  W_m^{\prime\, + } ) \cdot F( \delta  W_m^{\prime\, - } )  \,, \non
F(  \delta  W_m^{\prime\, + }  )  &=& \f{ m_{Z^\prime}}{ \sqrt{2} } \( \vec \nabla_\xi \cdot  \delta \vec W^{\prime\, + }   - \f{ i g_{Z^\prime } c_{ \vartheta_S }^2 }{ 2 }  \f{ \bar \zeta ( \xi ) }{ \xi }  ( s_\varphi \delta  W_{\xi_1}^{\prime\, + } - c_\varphi  \delta  W_{\xi_2}^{\prime\, + }  ) \right.   \non
&& \left. +i c_{\vartheta_S }  \sqrt{ \f{3}{2} } \sum_\omega \f{ V_\omega }{ v_{331}}  \bar f_\omega ( \xi ) e^{i \varphi } \delta \pi_{W^\prime \,, \omega }^+  \) \,, \non
F( \delta  W_m^{\prime\, - }  ) &=& \f{ m_{Z^\prime}}{ \sqrt{2} } \( \vec \nabla_\xi \cdot  \delta \vec W^{\prime\, - }   + \f{ i g_{Z^\prime } c_{ \vartheta_S }^2 }{ 2 }  \f{ \bar \zeta ( \xi ) }{ \xi }  (  s_\varphi \delta  W_{\xi_1}^{\prime\, - } - c_\varphi  \delta  W_{\xi_2}^{\prime\, - }  )  \right.  \non
&& \left.  - i c_{\vartheta_S }  \sqrt{ \f{3}{2} } \sum_\omega \f{ V_\omega }{ v_{331}}  \bar f_\omega ( \xi ) e^{ - i \varphi } \delta \pi_{ W^\prime \,, \omega }^- \)  \,.
\eeqn
Below, we transform the gauge fields to the polar coordinates by
\beqn
&& \left(  \ba{c}  \delta W_{\xi_1}^{\prime\, \pm }   \\[1mm]   \delta W_{\xi_2}^{\prime\, \pm } \ea \right) =  \left(  \ba{cc}  c_\varphi  &  - s_\varphi  \\ s_\varphi   & c_\varphi  \ea \right) \cdot \left(  \ba{c}   \delta W_\xi^{\prime\, \pm }   \\[1mm]   \delta W_\varphi^{\prime\, \pm }   \ea \right)   \,.
\eeqn

\para
The pure gauge perturbations in Eq.~\eqref{eq:gauge_perturb} are modified into
\beqn\label{eq:gauge_perturb_fixed}
%%%%%%%%%%%%%%%%%%%%%%%%%%%%%%%%%%%%%%%%%%%%%
\delta \tilde \mu_{W^\prime }&=&  \int d^2 \xi\, \Big\{ \f{ \partial \delta W_\xi^{ \prime\, + } }{ \partial \xi }  \f{ \partial \delta W_\xi^{ \prime\, - } }{ \partial \xi } +   \f{ \partial \delta W_\varphi^{ \prime\, + } }{ \partial \xi }  \f{ \partial \delta W_\varphi^{ \prime\, - } }{ \partial \xi }   \non
&& + \f{1}{ \xi^2} \( \delta W_\xi^{\prime\, +}  + \f{ \partial \delta W_\varphi^{\prime\, +}   }{ \partial \varphi }  \) \( \delta W_\xi^{\prime\, -}  + \f{ \partial \delta W_\varphi^{\prime\, -}  }{ \partial \varphi }  \)  +   \f{1}{ \xi^2} \(  \delta W_\varphi^{\prime\, +}  -  \f{ \partial \delta W_\xi^{\prime\, +}  }{ \partial \varphi }   \)  \(  \delta W_\varphi^{\prime\, -} -  \f{ \partial \delta W_\xi^{\prime\, -}  }{ \partial \varphi }   \)   \non
&&   + \f{g_{Z^\prime }^2  c_{\vartheta_S }^4 }{  4 }  \frac{ \bar \zeta^2 (\xi ) }{  \xi^2 } ( \delta  W_\xi^{ \prime\, + }  \delta W_\xi^{ \prime \, -} + \delta  W_\varphi^{ \prime\, + }  \delta W_\varphi^{ \prime \, -} ) + \frac{i g_{Z^\prime }  c_{\vartheta_S}^2  }{2 } \f{1}{ \xi} \f{ \partial \bar \zeta ( \xi)  }{ \partial \xi}    \(  \delta  W_\xi^{\prime\, +}  \delta W_\varphi^{\prime\, -} -  \delta W_\xi^{\prime\, - } \delta W_\varphi^{\prime\, +}  \)  \non
&& +  \f{ 3  c_{\vartheta_S }^2  }{2  } \( \sum_\omega \f{ V_\omega^2 }{ v_{331}^2 } \bar f_\omega^2 ( \xi ) \)  ( \delta  W_\xi^{ \prime\, + }  \delta W_\xi^{ \prime \, -}  + \delta  W_\varphi^{ \prime\, + }  \delta W_\varphi^{ \prime \, -}  )  \non
&&   + \f{ i g_{Z^\prime } c_{ \vartheta_S }^2 }{ 2 } \f{\bar \zeta( \xi ) }{ \xi^2 }  \[  -  \( \xi \f{ \partial W^{\prime\, + }_\xi }{\partial \xi } +  \delta W_\xi^{\prime\, + } + \f{ \partial W^{\prime\, + }_\varphi }{ \partial \varphi }  \)   \delta  W_\varphi^{\prime \, -}   + \(  \xi \f{ \partial W^{\prime\, - }_\xi }{\partial \xi } +  \delta W_\xi^{\prime\, - } +  \f{ \partial W^{\prime\, - }_\varphi }{ \partial \varphi }  \)  \delta  W_\varphi^{\prime\, + }  \right. \non 
&& \left.  +  \( \xi \f{\partial \delta W_\varphi^{\prime\, + } }{\partial \xi}+   \delta W_\varphi^{\prime\, + }  -  \f{ \partial  \delta W_\xi^{\prime\, +} }{ \partial \varphi }   \)  \delta  W_{ \xi }^{\prime\, - }  -  \( \xi \f{\partial \delta W_\varphi^{\prime\, - } }{\partial \xi}+   \delta W_\varphi^{\prime\, -}  -  \f{ \partial  \delta W_\xi^{\prime\, -} }{ \partial \varphi }   \) \delta  W_{ \xi }^{\prime\, + }     \]  \Big\}   \,.
\eeqn
The pure scalar perturbations in Eq.~\eqref{eq:scalar_perturb} are modified into
\beqn\label{eq:scalar_perturb_fixed}
%%%%%%%%%%%%%%%%%%%%%%%%%%%%%%%%%%%%%%%%%%%%%
\delta \tilde \mu_\pi &=&    \int d^2 \xi \, \Big\{  \sum_\omega \delta \pi_{W^\prime \,, \omega }^- \cdot  \(   - \f{ \partial^2 }{\partial \xi^2 } - \f{1}{ \xi } \f{ \partial }{ \partial \xi} - \f{1 }{ \xi^2 } \f{ \partial^2 }{  \partial \varphi^2 }  -  i g_{Z^\prime } (  \f{1}{ 3  } - s_{ \vartheta_S }^2 )   \f{\bar \zeta ( \xi ) }{ \xi^2 } \f{\partial }{ \partial \varphi }   \right. \non
&& \left.  + \f{g_{ Z^\prime }^2 }{4} (  \f{1}{ 3  } - s_{ \vartheta_S }^2 )^2  \f{\bar \zeta^2 ( \xi )}{ \xi^2 }  \) \delta \pi_{ W^\prime \,, \omega }^+   + \sum_\omega V(  \delta \pi_{W^\prime  \,, \omega }^\pm    )    \non
%&&  + \hf \delta \pi_{ W^\prime \,, \omega }^+ \cdot   \( - \f{ \partial^2 }{\partial \xi^2 } - \f{1}{ \xi } \f{ \partial }{ \partial \xi} - \f{1 }{ \xi^2 } \f{ \partial^2 }{  \partial \varphi^2 } +  i g_{Z^\prime } (  \f{1}{ 3  } - s_{ \vartheta_S }^2 )  \f{\bar \zeta ( \xi ) }{ \xi^2 } \f{\partial }{ \partial \varphi } + \f{g_{ Z^\prime }^2 }{4} (  \f{1}{ 3  } - s_{ \vartheta_S }^2 )^2  \f{\bar \zeta^2 ( \xi )}{ \xi^2 }  \) \delta \pi_{ W^\prime \,, \omega }^-  \non
%
&&  +  \f{  3 c_{\vartheta_S }^2}{ 2 } \(  \sum_{\omega_1 } \f{ V_{\omega_1} }{ v_{331}  } \bar f_{\omega_1} ( \xi )  \delta \pi_{W^\prime  \,, \omega_1 }^+ \) \(  \sum_{\omega_2 } \f{ V_{\omega_2 } }{ v_{331} } \bar f_{\omega_2 } ( \xi )  \delta \pi_{W^\prime  \,, \omega_2 }^- \)    \Big\}    \,.
\eeqn
Notice, the summation in the last line contains terms both for the $\omega_1=\omega_2$ cases and the $\omega_1\neq \omega_2$ case.
The perturbed couplings between scalars and the gauge fields in Eq.~\eqref{eq:scalar_gauge_perturb} are modified into
\beqn\label{eq:scalar_gauge_perturb_fixed}
%%%%%%%%%%%%%%%%%%%%%%%%%%%%%%%%%%%%%%%%%%%%%
\delta \tilde \mu_c &=&  \sqrt{ 6 } c_{\vartheta_S } \sum_\omega  \f{  V_\omega }{ v_{331}  }  \int d^2 \xi \,   \Big\{  i  \f{d \bar f_\omega (\xi ) }{d \xi}   \( e^{-i \varphi } \delta  W_\xi^{\prime \, +}  \delta \pi_{ W^\prime \,, \omega }^-    -  e^{ i \varphi }   \delta  W_\xi^{\prime \, -}   \delta \pi_{ W^\prime \,, \omega }^+   \)   \non
&& + \( 1 +   \f{   g_{Z^\prime }   }{ 3 } \bar \zeta ( \xi )  \)  \f{ \bar f_\omega ( \xi) }{ \xi }   \(  e^{-i \varphi }  \delta  W_\varphi^{\prime \, +}   \delta \pi_{ W^\prime \,, \omega }^-   + e^{ i \varphi }   \delta  W_\varphi^{\prime \, -}   \delta \pi_{ W^\prime \,, \omega }^+    \)   \Big\} \,.
\eeqn

\subsection{The stability matrix in the $Z^\prime$ string background}
\label{section:331_matrix}

\para
To analyze the stability, it is convenient to transform the perturbed gauge fields into the spin eigenstates as
\beqn
&& \delta W_{ \uparrow }^{\prime\, \pm }= \f{ e^{-i \varphi} }{ \sqrt{2} } ( \delta W_\xi^{ \prime\, \pm } - i \delta W_\varphi^{ \prime\, \pm } ) \,, \quad \delta W_{ \downarrow }^{\prime\, \pm }= \f{ e^{ i \varphi} }{ \sqrt{2} } ( \delta W_\xi^{ \prime\, \pm } + i \delta W_\varphi^{ \prime\, \pm } )  \,
%&&   \delta N_{ \uparrow } = \f{ e^{-i \varphi} }{ \sqrt{2} } ( \delta N_\xi - i \delta N_\varphi ) \,, \quad \delta N_{ \downarrow } = \f{ e^{ i \varphi} }{ \sqrt{2} } ( \delta N_\xi + i \delta N_\varphi ) \,.
\eeqn
such that $( \delta W_{ \uparrow }^{\prime\, \pm })^\dag =  \delta W_{ \downarrow }^{\prime\, \mp }$.
Below, we perform the Fourier expansion to the perturbed modes
\beqn
&& \delta \pi_{W^\prime\,, \omega}^+ = \sum_\ell s_{\ell,\omega} ( \xi) e^{- i\ell\varphi} \,, \quad  \delta \pi_{W^\prime\,, \omega}^-= \sum_\ell s_{\ell,\omega}^* (\xi)e^{ i\ell\varphi} \non
&& \delta W^{\prime\, +}_\uparrow= \sum_\ell -iw_{ \uparrow\,, \ell}  ( \xi) e^{ - i \ell \varphi} \,, \quad  \delta W^{\prime\, +}_\downarrow= \sum_\ell iw_{\downarrow \,,  \ell } ( \xi) e^{- i (  \ell - 2 )\varphi} \non
&& \delta W^{\prime\, -}_\uparrow = \sum_\ell -iw_{ \downarrow \,, \ell }^*( \xi) e^{i( \ell -2 )\varphi} \,, \quad  \delta W^{\prime\, -}_\downarrow= \sum_\ell iw_{ \uparrow \,,  \ell }^* ( \xi) e^{i \ell \varphi} \,.
\eeqn

\para
In the basis of $\delta \Theta \equiv (s_{\ell \,, 1} \,, s_{\ell \,, 2} \,, w_{\uparrow \,, \ell}, w_{\downarrow\,, \ell } )^T$, the perturbed string tension is expressed in terms of the stability matrix
\beqn\label{eq:331_matrix}
\tilde \mu &=&  \f{g_{Z^\prime }^2 }{9 }  \int \xi d \xi \,  \delta \Theta^{ \dag } \hat \Oc  \delta \Theta  \,, \non
&& \hat \Oc    =
\begin{pmatrix}
\mathcal{D}_{11} & \Dc_{12} & \Bc_{ \uparrow} &  \Bc_{ \downarrow} \\
\Dc_{12} &  \Dc_{22} & \Bc_{ \uparrow}   &  \Bc_{ \downarrow} \\
\Bc_{ \uparrow}  & \Bc_{ \uparrow} & \mathcal{D}_{ \uparrow} & 0\\
 \Bc_{ \downarrow} & \Bc_{ \downarrow}  &0  & \mathcal{D}_{ \downarrow } 
\end{pmatrix}   \,,
\eeqn
where the elements of the stability matrix of $\hat \Oc$ read
\beqs\label{eqs:331_matrix_stable}
\beqn
\mathcal{D}_{11}&=& - \frac{\partial^2}{\partial\xi^2} - \frac{1}{\xi}\frac{\partial}{\partial\xi} +\frac{1}{\xi^2} \( \ell+  g_{Z^\prime} (-\frac{1}{ 6 } + \hf s_{\vartheta_S}^2)\bar{\zeta} ( \xi )  \)^2 +\frac{3 c_{\vartheta_S}^2 }{2}  \sum_\omega \bar{f}_{\omega}^2 ( \xi ) \frac{ V^2_\omega}{ v_{331}^2} \non
&& + \beta_1 c_{\tilde \beta}^2 ( \bar f_1^2 ( \xi) -1) + \beta_3 \(   c_{\tilde \beta}^2 ( \bar f_1^2 ( \xi) -1) +  s_{\tilde \beta}^2 ( \bar f_2^2 ( \xi) -1) \) + \beta_4 s_{\tilde \beta}^2 \bar f_2^2 ( \xi )  \,, \label{eq:331_stable_D11}\\[2mm]
\mathcal{D}_{22}&=& - \frac{\partial^2}{\partial\xi^2} - \frac{1}{\xi}\frac{\partial}{\partial\xi} +\frac{1}{\xi^2} \( \ell+  g_{Z^\prime} (-\frac{1}{ 6 } + \hf s_{\vartheta_S}^2)\bar{\zeta} ( \xi ) \)^2 +\frac{3 c_{\vartheta_S}^2 }{2}  \sum_\omega \bar{f}_{\omega}^2 ( \xi ) \frac{ V^2_\omega}{ v_{331}^2} \non 
&& + \beta_2 s_{\tilde \beta}^2 ( \bar f_2^2 ( \xi) -1) +   \beta_3 \(   c_{\tilde \beta}^2 ( \bar f_1^2 ( \xi) -1) +  s_{\tilde \beta}^2 ( \bar f_2^2 ( \xi) -1) \) + \beta_4 c_{\tilde \beta}^2 \bar f_1^2 ( \xi ) \,,\label{eq:331_stable_D22} \\[2mm]
\mathcal{D}_{12}&=& \[  - \beta_4    \bar f_1( \xi ) \bar f_2 ( \xi )   + \beta_5  \( \bar f_1( \xi ) \bar f_2 ( \xi ) -  1  \) + \f{3 c_{\vartheta_S }^2 }{2 }  \bar f_1( \xi ) \bar f_2 ( \xi ) \]  s_{\tilde \beta} c_{\tilde \beta}  \,,\label{eq:331_stable_D12} \\[2mm]
\Bc_{ \uparrow}&=& \sqrt{3 } c_{\vartheta_S} \sum_\omega \frac{ V_\omega}{ v_{331} }\left[  \frac{\partial \bar{f}_\omega}{\partial \xi}+\frac{ \bar{f}_\omega( \xi ) }{\xi} \( 1 +\frac{g_{Z^\prime }}{3} \bar{\zeta} ( \xi ) \) \right] \,, \\[2mm]
\Bc_{ \downarrow}&=& \sqrt{ 3 }c_{\vartheta_S} \sum_\omega \frac{ V_\omega}{\ v_{331} }\left[ -\frac{\partial \bar{f}_\omega}{\partial \xi}+\frac{ \bar{f}_\omega( \xi ) }{\xi} \( 1 +\frac{g_{Z^\prime }}{ 3 }\bar{\zeta} (\xi) \) \right] \,, \\[2mm]
\mathcal{D}_{ \uparrow} &=& -\frac{\partial^2}{\partial\xi^2}- \frac{1}{\xi}\frac{\partial}{\partial\xi}+\frac{1}{\xi^2} \( \ell - \frac{ g_{Z^\prime} }{ 2 } c_{\vartheta_S}^2\bar{\zeta} ( \xi) \)^2  \non 
&& + g_{Z^\prime } c_{\vartheta_S}^2\frac{1}{\xi}\frac{\partial \bar{\zeta} ( \xi ) }{\partial \xi}+\frac{ 3 c_{\vartheta_S}^2 }{2 } \sum_\omega  \bar{f}_{\omega}^2 ( \xi ) \frac{ V^2_\omega}{ v_{331}^2}\,,\label{eq:331_matrix_Dup} \\[2mm]
\mathcal{D}_{ \downarrow }&=& - \frac{\partial^2}{\partial\xi^2} - \frac{1}{\xi}\frac{\partial}{\partial\xi}+\frac{1}{\xi^2} \( \ell - 2 - \frac{ g_{Z^\prime }  }{2 } c_{\vartheta_S}^2\bar{\zeta} ( \xi ) \)^2 \non
&&  - g_{Z^\prime } c_{\vartheta_S}^2\frac{1}{\xi}\frac{\partial \bar{\zeta} ( \xi ) }{\partial \xi}+\frac{3 c_{\vartheta_S}^2}{ 2}  \sum_\omega \bar{f}_{\omega}^2 ( \xi ) \frac{ V^2_\omega}{v_{331}^2} \,. \label{eq:331_matrix_Ddown}
\eeqn
\eeqs
Several features of the stability matrix are the following.
\begin{enumerate}

\item In the semilocal limit of $\vartheta_S\to \f{\pi}{2}$ ($c_{\vartheta_S}\to 0$), there can still be off-diagonal elements of $\Dc_{12}$, which are due to the mutual self couplings between two Higgs fields.
Meanwhile in the SM, the stability matrix becomes diagonal in its semilocal limit of $\vartheta_W\to \f{\pi}{2}$ ($c_{\vartheta_W}\to 0$)~\cite{Eto:2024xvc}.

\item The Landau levels of $\ell$ only appear in the diagonal elements.

\item For diagonal elements of $( \Dc_{11} \,, \Dc_{22}\,, \Dc_\uparrow\,, \Dc_{\downarrow})$, any term besides of the two-dimensional Laplacian operator can potentially destabilize the $Z^\prime$ string when they are negative.

\item The off-diagonal element of $\Dc_{12}$ are due to the mixings between $\delta \pi_{W^\prime\,, 1}^\pm$ and $\delta \pi_{W^\prime\,, 2}^\mp$ from the perturbed Higgs potential in Eq.~\eqref{eq:331potential_perturb} as well as the gauge fixing term in Eq.~\eqref{eq:331_fix}. 
They are absent if we only assumed one ${\rm SU}(3)_W \otimes {\rm U}(1)_X$ anti-fundamental Higgs field.

\end{enumerate}

\para
In the SM, one major source of the $Z$ string instability is due to the $W$ condensate~\cite{Ambjorn:1989sz,Vachaspati:1992jk,Achucarro:1993bu,Perkins:1993qz}.
Similarly, we also expect the $Z^\prime$ string instability when the corresponding magnetic fields are sufficiently strong.
In the 331 model and more generic ${\rm SU}(N)\otimes {\rm U}(1)_X$ models, we dub this as the {\it off-diagonal gauge boson condensate}.
With the $Z^\prime$ string profile in Eq.~\eqref{eq:331_string_form}, the corresponding magnetic field is $\vec B_{Z^\prime} = B_{Z^\prime} \vec e_z$ and it couples the off-diagonal 331 gauge bosons of $(W_\mu^{\prime\, \pm}\,, N_\mu\,, \bar N_\mu)$ with strength of $g_{Z^\prime } c_{ \vartheta_S }^2$ according to Eq.~\eqref{eq:gauge_perturb_fixed} and the $\Dc_{\uparrow\,, \downarrow}$ terms in Eqs.~\eqref{eq:331_matrix_Dup} and \eqref{eq:331_matrix_Ddown}.
The energy dispersion of the charged $W^{\prime\, \pm}$ is described by the Landau levels in the $(x\,,y)$-plane of
\beqn
E^2&=& ( 2\ell + 1 - 2 S_z) g_{Z^\prime } c_{ \vartheta_S }^2 B_{Z^\prime} + p_z^2 + m_{W^\prime }^2 \,.
\eeqn
With $\ell=1$ and $S_z= +1$, the energy of charged $W^{\prime\, \pm}$ will become negative when the magnetic field is stronger than
\beqn
&& B_{Z^\prime} > \f{ m_{Z^\prime}^2 }{g_{Z^\prime} c_{ \vartheta_S}^2 } = \f{3}{4} \f{ m_{Z^\prime}^2 }{g_{Z^\prime} } \,,
\eeqn
where we used the gauge boson masses squared in Eq.~\eqref{eq:331_GBmass}.
The same results also hold for the off-diagonal $(N_\mu\,, \bar N_\mu)$ gauge bosons.
The instability here is due to the spin magnetic term of $g_{Z^\prime } c_{\vartheta_S}^2\frac{1}{\xi}\frac{\partial \bar{\zeta} ( \xi ) }{\partial \xi}$ in the $\Dc_\uparrow$ element.
This is due to our convention of the $Z^\prime$ string profile in Eq.~\eqref{eq:331_string_form}, with a negative derivative of $\f{\partial \bar \zeta( \xi)}{\partial \xi}$ according to the string profiles displayed in Fig.~\ref{fig:331_profiles}.

\para
The second source of the $Z$ string instability in the SM is due to heavy mass of the SM Higgs boson.
In the context of the 331 model, one expects the similar source by focusing on the $( \Dc_{11}\,, \Dc_{22})$ terms and turning off the self couplings of $(\beta_3\,, \beta_4 \,, \beta_5)$ in Eqs.~\eqref{eq:331_stable_D11} and \eqref{eq:331_stable_D22}.
The increases of the positive values of $(\beta_1\,, \beta_2)$ tends to destabilize the scalar perturbed modes, with the negative contributions of $\bar f_{1\,,2}^2 (\xi) -1$ when one approaches to the string core.
On the other hand, the mutual self couplings of $\beta_3$ is possible to cancel the instability from the positive $(\beta_1\,, \beta_2)$ when one sets it to be reasonably negative.
For the simplified case of $t_{\tilde \beta}=1$ and $\lambda_1=\lambda_2=\lambda_5$ discussed in Sec.~\ref{section:string_sol}, this means the relations of $  \beta_3> - \hf ( \beta_{1 \,, 2} + \beta_5 )$ should be satisfied.

%###################################################################
\vspace*{3mm}
\section{Numerical results}
\label{section:numerical}
%###################################################################

\para
The numerical analysis the string stability relies on the solutions of the following eigenvalue equation
\beqn
&& \hat \Oc \delta \Theta = \omega^2 \delta \Theta\,,
\eeqn
where a negative eigenvalue of $\omega^2$ signifies the unstable region.
Below, we will focus on the $s$-wave solution with $\ell=0$ in Eqs.~\eqref{eq:331_matrix} and \eqref{eqs:331_matrix_stable}.
The numerical code for this section can be found in \cite{BianZY}.
%The numerical code for this section can be found in \cite{Bian:331stability}.

\subsection{The critical point in the semilocal limit of $c_{\vartheta_S}=0$}
\label{section:331_critical}

 %%%%%%%%%%%%%%%%%%%%%%%%%%%%%%%%%%%%%%%%%%%%%%%%%%%%%%%
\begin{figure}[htb]
\centering
\includegraphics[height=5.4cm]{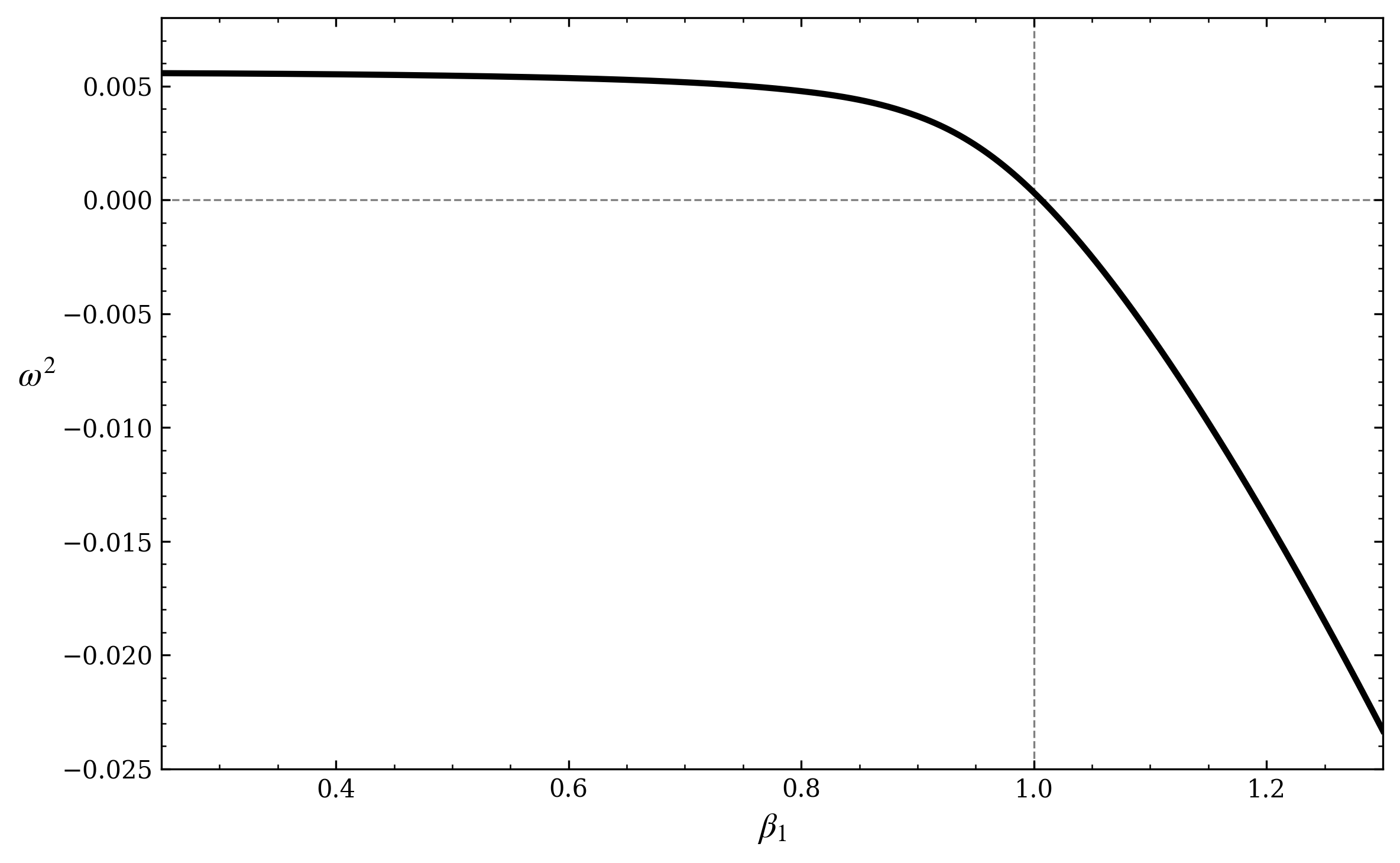}
\caption{The dependence of the eigenvalue $\omega^2$ on $\beta_1$ for the one 331 Higgs triplet case by setting $\bar f_1(\xi)\neq 0$, $\bar f_2(\xi)=0$, and $c_{\vartheta_S}=0$.}
\label{fig:beta1_f1critical} 
\end{figure}
%%%%%%%%%%%%%%%%%%%%%%%%%%%%%%%%%%%%%%%%%%%%%%%%%%%%%%%

\para
First, we perform a consistent check of the numerical calculation by considering only one non-vanishing Higgs field profile of $\bar f_1(\xi)$, and setting $\bar f_2(\xi)=0$ by hand in Eqs.~\eqref{eqs:331string_EOM_dimensionless}.
This also means that $(c_{\tilde \beta} \,, s_{\tilde \beta})=(1\,,0)$ and there is only one non-vanishing parameter of $\beta_1$ in the Higgs potential.
We also take the semilocal limit of $c_{\vartheta_S}=0$, which is similar to the semilocal limit of $c_{\vartheta_W}=0$ that was previously found in the SM.
Correspondingly, the stability matrix in Eq.~\eqref{eq:331_matrix} are simplified into a diagonal matrix with the following elements
\beqs
\beqn
&& \mathcal{D}_{11} = - \frac{\partial^2}{\partial\xi^2} - \frac{1}{\xi}\frac{\partial}{\partial\xi} +\frac{1}{\xi^2} \( \ell+  \frac{ g_{Z^\prime} }{ 3 }  \bar{\zeta} ( \xi )  \)^2   + \beta_1  ( \bar f_1^2 ( \xi) -1) \,, \\[2mm]
%%
%&& \mathcal{D}_{22} = - \frac{\partial^2}{\partial\xi^2} - \frac{1}{\xi}\frac{\partial}{\partial\xi} +\frac{1}{\xi^2} \( \ell+   \frac{ g_{Z^\prime}}{ 3 }  \bar{\zeta} ( \xi ) \)^2  \,, \\[2mm]
%%
&&  \Bc_{ \uparrow} = \Bc_{ \downarrow} = 0\,, \\[2mm]
&& \mathcal{D}_{ \uparrow} = -\frac{\partial^2}{\partial\xi^2}- \frac{1}{\xi}\frac{\partial}{\partial\xi}+\frac{ \ell^2 }{\xi^2}  \,, \\[2mm]
&& \mathcal{D}_{ \downarrow }= - \frac{\partial^2}{\partial\xi^2} - \frac{1}{\xi}\frac{\partial}{\partial\xi}+\frac{( \ell-2)^2 }{\xi^2}  \,.
\eeqn
\eeqs
As was recently pointed out in Ref.~\cite{Eto:2024xvc}, the Helmholtz equations from the simplified operators of $( \Dc_\uparrow\,, \Dc_\downarrow)$ must yield positive eigenvalues.
Thus the only instability is due to the operator of $\Dc_{11}$, since the last term of $\beta_1( \bar f_1^2(\xi) -1)$ is negative when one approaches to the string core according to the upper panels in Fig.~\ref{fig:331_profiles}~\footnote{Notice that $\beta_1>0$, since the Higgs potential should be bounded from below.}.
The eigenvalue of the differential operator $\Dc_{11}$ with the varying parameter $\beta_1$ is displayed in Fig.~\ref{fig:beta1_f1critical}.
Indeed, the critical point between the stable and unstable region sits at $\beta_{1\,,c}=1.0$, which was identical to what was previously known for the semilocal string~\cite{Vachaspati:1991dz,Hindmarsh:1991jq}.
More generally, the critical point of $\beta_{1\,,c}=1.0$ with one Higgs multiplet is universal for any ${\rm SU}(N) \otimes {\rm U}(1)_X\to {\rm SU}(N-1) \otimes {\rm U}(1)_{X^\prime}$ breaking.
We show this point by obtaining the self-dual equations by a gauge-covariant approach proposed by Bogomol'nyi~\cite{Bogomolny:1975de} in Appendix~\ref{section:SDeq}.

\subsection{The detailed analysis of the instability}
\label{section:331_details}

\para
Next, we proceed to present the 331 string stability for more generic parameter inputs.
The individual contributions from the stability matrix elements in Eqs.~\eqref{eqs:331_matrix_stable} will be analyzed in details.

%%%%%%%%%%%%%%%%%%%%%%%%%%%%%%%%%%%%%%%%%%%%%%%%%%%%%%%
\begin{figure}[htb]
\centering
\includegraphics[height=5.4cm]{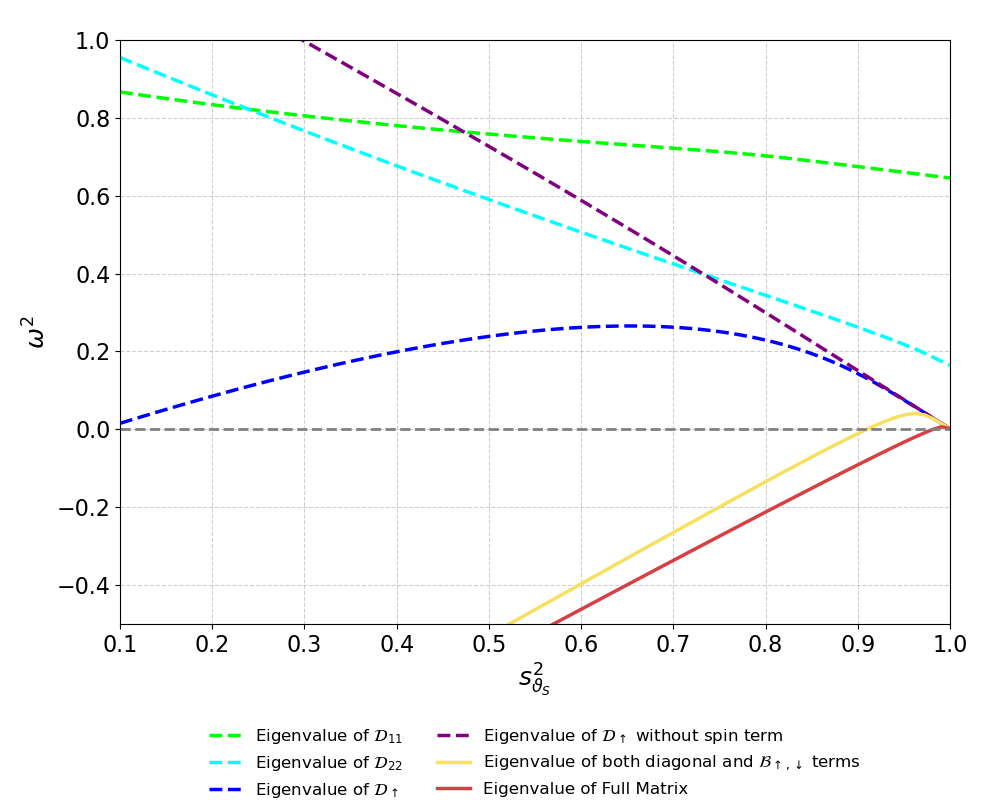}
\includegraphics[height=5.4cm]{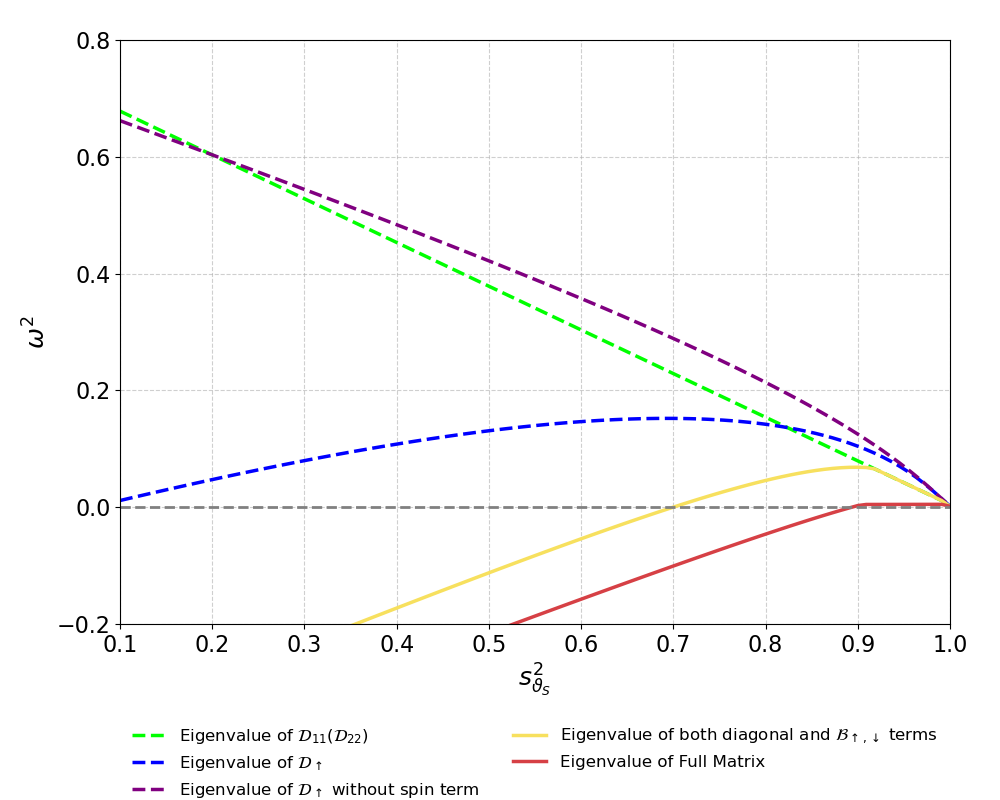}
\caption{The eigenvalues $\omega^2$ of stability matrix elements with $\lambda_3=-0.05$ (left panel) and $\lambda_3=- 0.145$ (right panel) versus $s_{\vartheta_S}^2$, and $g_{Z^\prime}=1.0592$.
Parameters in the left panel are: $\lambda_1=\lambda_2=0.15$, $\lambda_4=\lambda_5=0.1$, and $t_{\tilde \beta}=2$.
Parameters in the right panel are: $\lambda_1=\lambda_2=\lambda_5=0.15$, $\lambda_4=0$, and $t_{\tilde \beta}=1$.
}
\label{fig:eigenvalue} 
\end{figure}
%%%%%%%%%%%%%%%%%%%%%%%%%%%%%%%%%%%%%%%%%%%%%%%%%%%%%%%

\para
In Fig.~\ref{fig:eigenvalue}, we demonstrate the eigenvalues $\omega^2$ of different stability matrix elements versus the 331 mixing angle $s_{\vartheta_S}^2$, with two different negative inputs of $\lambda_3$.
The left (right) panels take the same parameter choices in the left (right) panels of Fig.~\ref{fig:331_profiles}.
In both panels, the eigenvalues of the diagonal elements $(\Dc_{11}\,, \Dc_{22})$ are positive while decreasing as $\vartheta_S\to \f{\pi}{2}$.
This is because the positive contributions from the gauge fixing terms are $\propto c_{\vartheta_S}^2$, while the terms from the Higgs potential that are $\propto \beta_{1\,,2\,,3}$ always contribute negatively.
The $\Dc_\uparrow$ elements contain the spin magnetic term, and the inclusion of this term significantly reduce the corresponding eigenvalues, as one compares the dashed purple curves and dashed blue curves.
By further inclusion of the off-diagonal $\Bc_{\uparrow\,, \downarrow}$ elements, the joint effects of $\Dc_\uparrow$ and $\Bc_{\uparrow\,, \downarrow}$ have already point to the semilocal limits in both panels.

%%%%%%%%%%%%%%%%%%%%%%%%%%%%%%%%%%%%%%%%%%%%%%%%%%%%%%%
\begin{figure}[htb]
\centering
\includegraphics[height=5.4cm]{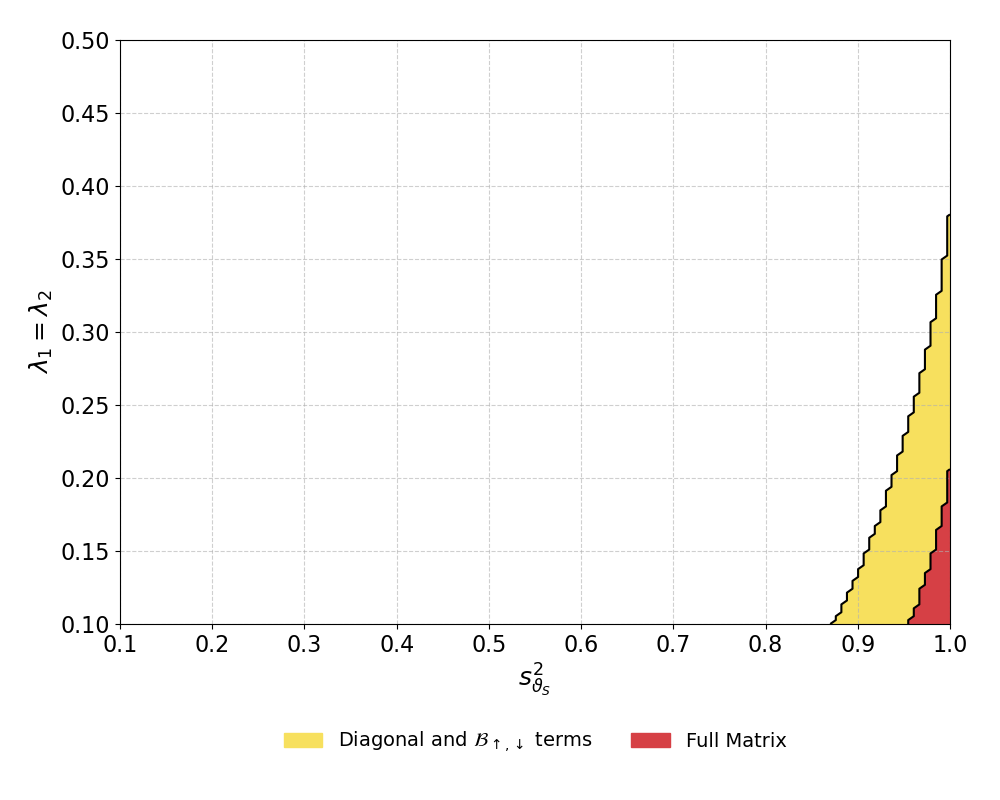}
\includegraphics[height=5.4cm]{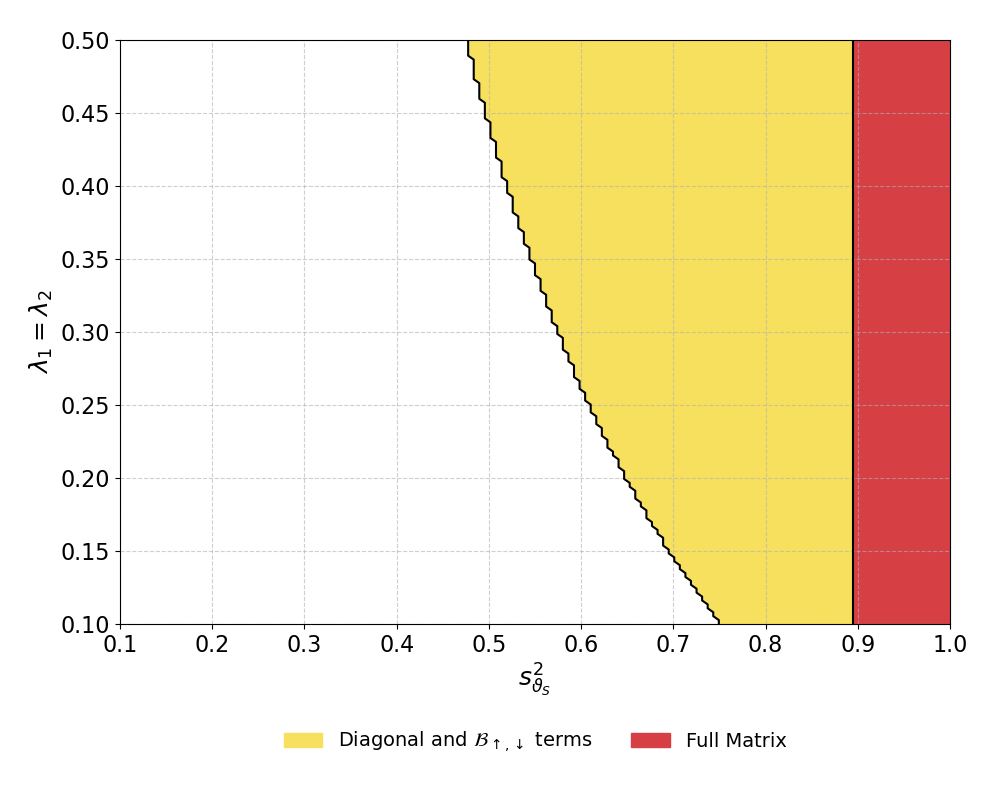}
\caption{The stability regions with a fixed $\lambda_3=-0.05$ (left panel) and a varying $\lambda_3=- \hf ( \lambda_1 + \lambda_5) +0.005$ (right panel) in the $( s_{\vartheta_S}^2 \,, \lambda_1=\lambda_2)$ plane, with $g_{Z^\prime}=1.0592$.
Other parameters fixed for all plots are $\lambda_4=\lambda_5=0.1$, $t_{\tilde \beta}=2$ in the left panel, and $\lambda_1= \lambda_2=\lambda_5$, $\lambda_4=0$, $t_{\tilde \beta}=1$ in the right panel.}
\label{fig:stability_region} 
\end{figure}
%%%%%%%%%%%%%%%%%%%%%%%%%%%%%%%%%%%%%%%%%%%%%%%%%%%%%%%

\para
With the decompositions of the individual contributions to the eigenvalues, we further demonstrate the stability regions in Fig.~\ref{fig:stability_region}.
In the left panel, we fix the parameters of $\lambda_3=-0.05$ and $\lambda_4=\lambda_5=0.1$.
In the right panel, we vary the negative parameter as $\lambda_3=- \hf ( \lambda_1 + \lambda_5) +0.005$ (together with $\lambda_1=\lambda_2=\lambda_5$) in order to minimize the instability effects in the diagonal $(\Dc_{11}\,, \Dc_{22})$ elements due to the 331 Higgs potential.
The joint contributions from the diagonal elements of $( \Dc_{11}\,, \Dc_{22}\,, \Dc_{\uparrow\,, \downarrow})$ plus the off-diagonal elements of $\Bc_{\uparrow\,, \downarrow}$ are shown in the shaded yellow regions, and the completely stable region with all matrix elements are shown in the shaded red regions.
Since we fix the negative parameter of $\lambda_3=-0.05$ in the left panel, the increasings of $\lambda_1=\lambda_2$ enhance the instability sources from the diagonal elements of $(\Dc_{11}\,, \Dc_{22})$ due to the 331 Higgs potential.
With the special choice of $\lambda_3=- \hf ( \lambda_1 + \lambda_5) +0.005$ in the right panel, the instability source due to the perturbations to the 331 Higgs potential has been effectively erased.
Thus, the stable region only depends on the 331 mixing angle of $\vartheta_S$.
In both panels, the 331 $Z^\prime$ string can be stable when one approaches to the semilocal limit of $\vartheta_S\to \f{ \pi}{2}$.
Specifically, one has $s_{\vartheta_S}^2 \gtrsim 0.95$ in the left panel with $\lambda_1=\lambda_2=0.1$, and $s_{\vartheta_S}^2 \gtrsim 0.9$ in the right panel for all $\lambda_1=\lambda_2$.
By using the definition in Eq.~\eqref{eq:Salam_angle}, this means the 331 gauge couplings should satisfy $\f{g_X}{g_{3W}}\gtrsim 7.55$ for the left panel, or $\f{g_X}{g_{3W}}\gtrsim 5.20$ for the right panel.

%###################################################################
\vspace*{3mm}
\section{Conclusion}
\label{section:conclusion}
%###################################################################

\para
In this paper, we carry out the detailed studies of the non-topological $Z^\prime$ string in the minimal 331 model, where we consider this as an effective theory after the GUT-scale breaking of a toy ${\rm SU}(6)$ model.
The corresponding global symmetries in the ${\rm SU}(6)$ fermion sector in Eq.~\eqref{eq:SU6_flavor} suggests two 331 anti-fundamental Higgs fields for the sequential symmetry breaking pattern of ${\rm SU}(3)_c \otimes {\rm SU}(3)_W \otimes {\rm U}(1)_X\to {\rm SU}(3)_c \otimes {\rm SU}(2)_W \otimes {\rm U}(1)_Y$.

\para
The non-topological $Z^\prime$ string carries the quantized magnetic flux in Eq.~\eqref{eq:331_string_flux}.
The string profiles and the variation of the string energy densities within and outside of the core are obtained numerically and displayed in Fig.~\ref{fig:331_profiles}.
The stability analysis largely follows the methods in the $Z$ string of the electroweak sector, where one includes the perturbations to the off-diagonal massive gauge bosons as well as the scalar fields that contain the corresponding Nambu-Goldstone bosons.
For the 331 model, it is sufficient to analyze the quadratic terms of the perturbed string tension with the $(\delta W_m^{\prime\, \pm}\,, \delta \pi_{W^\prime\,, \omega}^\pm)$.
By including the gauge fixing terms and performing the Fourier expansions, we obtain the stability matrix in Eqs.~\eqref{eq:331_matrix} and \eqref{eqs:331_matrix_stable}.
The detailed analysis of the full stability matrix with different 331 parameters show that the $Z^\prime$ string can only be stable when one approaches to the semilocal limit of $\vartheta_S\to \f{\pi}{2}$.
By converting the lower limits to the 331 mixing angles to the lower limits to the 331 gauge couplings in Fig.~\ref{fig:stability_region}, we find that $\f{g_X}{g_{3W}}\gtrsim 7.55$ for the left panel, or $\f{g_X}{g_{3W}}\gtrsim 5.20$ for the right panel.
If we convert the ${\rm U}(1)_X$ gauge coupling into the ${\rm U}(1)_1$ gauge coupling of the toy ${\rm SU}(6)$ model by $g_1= \f{2}{ \sqrt{3} } g_X$~\cite{Chen:2021haa}, the lower limits to the 331 mixing angles lead to $\f{g_1}{g_{3W}}\gtrsim 8.7$ for the left panel, or $\f{g_1}{g_{3W}}\gtrsim 6$ for the right panel.
These results are incompatible with the unification relations, either in the conventional ${\rm SU}(6)$ Lie group where $g_{3W}(v_U)=g_1(v_U)$ is required, or in the affine $\widehat{\rm SU}(6)_{k=1}$ Lie group where $2g_{3W}(v_U)=g_1(v_U)$ is required~\cite{Chen:2024wcj}.
One can further expect such semilocal limit to the mixing angles in the extended gauge sectors can be expected in general. 
Therefore, the non-topological $Z^\prime$ strings with even larger ${\rm SU}(N)\otimes {\rm U}(1)$ Lie groups are not likely to be classically stable.
Nevertheless, the multiple Higgs fields in the class of extended ${\rm SU}(N)$ theories are generally expected.
One can further look for the topological strings with the spontaneously broken global $\widetilde{\rm U}(1)$ symmetries, and this was previously discussed in Refs.~\cite{Dvali:1993sg,Eto:2018tnk,Eto:2021dca} for the two Higgs doublet models in the SM.

%###################################################################
\vspace*{3mm}
\section*{Acknowledgments}
%###################################################################
%
%
\para
N.C. thanks Shandong University for hospitality when preparing this work.
This work is partially supported by the National Natural Science Foundation of China (under Grant No. 12275140, No. 12475094, No. 12135006, and No. 12575099) and Nankai University.

\appendix

%###################################################################
\vspace*{3mm}
\section{The self-dual equations for the ${\rm SU}(N) \otimes {\rm U}(1)_X\to {\rm SU}(N-1) \otimes {\rm U}(1)_{X^\prime}$ breaking}
\label{section:SDeq}
%###################################################################

\para
In this section, we will look for the self-dual equations in the extended weak sector by following a gauge-covariant approach proposed by Bogomol'nyi~\cite{Bogomolny:1975de}.
Such self-dual equations in the EW sector of the SM were previously analyzed in Refs.~\cite{Ambjorn:1988tm,Ambjorn:1989bd,Ambjorn:1989sz,Bimonte:1994ik}.
We start from the string tension of
\beqn\label{eq:SUN_tension}
\mu_{\rm ANO}&=& \int  d^2 x \, \Big[ \frac{1}{2 } ( \Wc_{ 12}^I )^2 +   \frac{1}{2} ( \Xc_{ 12 } )^2  + | D_m \Phi_{  \repb{N} } |^2 + V(  \Phi_{\repb{ N} })   \Big] \,, \non
&& V(  \Phi_{\repb{ N} })  =  \f{ \lambda_1 }{2} \( |\Phi_{  \repb{N} }|^2 - \hf \bar V^2 \)^2 \,,
\eeqn
where we denoted the ${\rm SU}(N) \otimes {\rm U}(1)_X$ field strength tensors as 
\beqn
&& \Wc_{12}^I  \equiv \partial_{ [1} \Wc_{2]}^I + g_N f^{ I J K } \Wc_1^J \Wc_2^K \,, ~~(I \,, J \,, K = 1\,, ... \,, N^2-1 ) \,, \quad \Xc_{12} \equiv \partial_{ [1} \Xc_{2]} \,,
\eeqn
and gauge couplings as $(g_N\,, g_X)$. 
The massive and massless $\uG(1)_Q$-neutral gauge bosons in terms of an ${\rm SU}( N)$ mixing angle as follows
\beqn\label{eq:SUN_mixing}
&&   \( \ba{c} \Wc_\mu^{N^2-1} \\  \Xc_\mu \ea \) = \( \ba{cc} c_{\vartheta_N} & s_{\vartheta_N} \\ -s_{\vartheta_N} & c_{\vartheta_N} \ea \) \( \ba{c} \Zc_\mu^\prime  \\  \Xc_\mu^\prime \ea \) \,, \non
&&  t_{\vartheta_N} = \f{g_X}{ c_N g_N}\,, \quad \text{with} \quad  c_N = \sqrt{\hf N(N-1)}  \,.
\eeqn
It is convenient to define the coupling for the massive $\Zc^\prime$ gauge boson as follows
\beqn\label{eq:SUN_gZp}
g_{Z^\prime }^2 &\equiv& c_N^2  g_N^2 + g_X^2  \,.
\eeqn
There is only one ${\rm SU}(N)$ anti-fundamental Higgs field of $\Phi_{  \repb{N} }$ assumed in the spectrum.

\para
One can rewrite the Higgs kinematic terms with the following identity
\beqn
|  D_m \Phi_{  \repb{N} }  |^2 %&=& \( (  D_1 + i D_2 ) \Phi_{  \repb{N} } \)^\dag  \( (  D_1 + i D_2 ) \Phi_{  \repb{N} }  \) - i \[ ( D_1 \Phi_{  \repb{N} } )^\dag  ( D_2 \Phi_{  \repb{N} } ) - ( D_2 \Phi_{  \repb{N} } )^\dag  ( D_1 \Phi_{  \repb{N} } )   \]  \non
&=& | (  D_1 + i D_2 ) \Phi_{  \repb{N} } |^2 + i \Phi_{  \repb{N} }^\dag \[ D_1  \,, D_2 \]  \Phi_{  \repb{N} }  - i \epsilon_{ mn} \partial_m J_n   \,, \non
{\rm with}&~& J_n \equiv \Phi_{  \repb{N} }^\dag D_n \Phi_{  \repb{N} }  \,, \quad  \[ D_1  \,, D_2 \] = + i g_N \Wc_{12}^I ( T^I )^T  +  i  \f{  g_X  }{N} \Xc_{12} \,.
\eeqn
The last term is a total derivative and we drop it below.
One can further express the string tension as follows
\beqn
\mu_{\rm ANO}&=&  \int d^2 x \,  \Big\{   \f{1}{2 }  \(  \Wc_{ 12 }^{I} -   g_N \Phi_{  \repb{N} }^\dag ( T^I )^T \Phi_{  \repb{N} }  \)^2  \non
&&  + \f{1}{2 } \Big(  \Xc_{12 } - \f{ g_X }{ N } | \Phi_{  \repb{N} } |^2  + \f{N }{ g_X } \( \f{( N-1) g_N^2}{ 4N }  + \f{ g_X^2}{ 2 N^2 }  \) \bar V^2 \Big)^2    + | (  D_1 + i D_2 ) \Phi_{  \repb{N} } |^2   \non
&&  +  \hf \(  \lambda_1  - \f{ ( N-1) g_N^2 }{2N }  - \f{ g_X^2 }{ N^2}  \) \( | \Phi_{  \repb{N} }  |^2 - \hf  \bar V^2  \)^2 \non
&&  - \f{ N }{ g_X } \(  \frac{ ( N-1) g_N^2 }{ 4N } + \f{ g_X^2 }{ 2N^2 }  \) \bar V^2 \Xc_{12}  \non
&&  - \( \f{ (N-1) g_N^2 }{4N} + \f{ g_X^2 }{ 2N^2 }  \) \(  \f{1}{ 4 } + \f{ N^2 }{2 g_N^2} ( \f{ (N-1) g_N^2 }{4N} + \f{ g_X^2 }{ 2N^2 } )  \) \bar V^4  \Big\}   \,,
\eeqn
where we have used the following relation
\beqn
\( \Phi_{  \repb{N} }^\dag (T^I )^T \Phi_{  \repb{N} }  \)^2  &=& \frac{ N - 1}{ 2N } \( \Phi_{  \repb{N} }^\dag \Phi_{  \repb{N} } \)^2 \,,
\eeqn
for the ${\rm SU}(N)$ Lie algebra.
The string tension is thus bounded from below as
\beqn
\mu_{\rm ANO}&\geq& -  \int d^2 x\,    \Big\{  \f{ g_{Z^\prime } }{ 2N s_{\vartheta_N} } \bar V^2 \Xc_{12}  + \f{ g_{Z^\prime }^2}{ 2N^2 } \(  \f{1}{ 4 } +  \f{ N (N-1) }{ 8 c_{\vartheta_N } }  \) \bar V^4  \Big\}  \,,
\eeqn
and the Bogmol'nyi bound is saturated with the following first-order equations and conditions
\beqs\label{eqs:SUN_SDeq}
\beqn
&& \Wc_{ 12 }^{I} =   \f{ g_{Z^\prime }}{ c_N} c_{\vartheta_N }   \Phi_{\repb{ N} }^\dag ( T^I )^T \Phi_{\repb{ N} }   \,,\label{eq:SUN_SD01}\\[2mm]
%%%%%%%%%%%%%%%%%%%%%%%%%%%%%%%%%%%%%%%%%%%%%
&& \Xc_{12 } = \f{ g_{Z^\prime } }{ N s_{\vartheta_N } }   \(  s_{\vartheta_N }^2  | \Phi_{\repb{ N} } |^2  -  \f{ 1 }{ 2  } \bar V^2 \)  \,,\label{eq:SUN_SD02}\\[2mm]
%%%%%%%%%%%%%%%%%%%%%%%%%%%%%%%%%%%%%%%%%%%%%
&& ( D_1 + i D_2 ) \Phi_{\repb{ N} } =0 \,,\label{eq:SUN_SD03}\\[2mm]
%%%%%%%%%%%%%%%%%%%%%%%%%%%%%%%%%%%%%%%%%%%%%
&&  \lambda_1 =   \f{ g_{Z^\prime }^2}{ N^2 }\,, \label{eq:SUN_SD04}
\eeqn
\eeqs
where we have used the relations in Eqs.~\eqref{eq:SUN_mixing} and \eqref{eq:SUN_gZp}.
The last condition in Eq.~\eqref{eq:SUN_SD04} suggest a definition of
\beqn\label{eq:SUN_beta1}
&& \beta_1 \equiv \f{N^2 \lambda_1}{ g_{Z^\prime }^2 } \,,
\eeqn
and this is consistent with the $\beta_i$ defined in Eq.~\eqref{eq:331_betai} for the 331 model.

%\bibliographystyle{utphys.bst}
%\bibliography{references}

\begin{thebibliography}{10}

\bibitem{Kibble:1976sj}
T.~W.~B. Kibble, ``{Topology of Cosmic Domains and Strings},''
  \href{http://dx.doi.org/10.1088/0305-4470/9/8/029}{{\em J. Phys. A}
  {\bfseries 9} (1976) 1387--1398}.

\bibitem{Zurek:1985qw}
W.~H. Zurek, ``{Cosmological Experiments in Superfluid Helium?},''
  \href{http://dx.doi.org/10.1038/317505a0}{{\em Nature} {\bfseries 317} (1985)
  505--508}.

\bibitem{Abrikosov:1956sx}
A.~A. Abrikosov, ``{On the Magnetic Properties of Superconductors of the Second
  Group},'' {\em Sov. Phys. JETP} {\bfseries 5} (1957) 1174--1182.

\bibitem{Nielsen:1973cs}
H.~B. Nielsen and P.~Olesen, ``{Vortex Line Models for Dual Strings},''
  \href{http://dx.doi.org/10.1016/0550-3213(73)90350-7}{{\em Nucl. Phys.}
  {\bfseries B61} (1973) 45--61}.
[,302(1973)].
%%CITATION = NUPHA,B61,45;%%.

\bibitem{Nambu:1977ag}
Y.~Nambu, ``{String-Like Configurations in the Weinberg-Salam Theory},''
  \href{http://dx.doi.org/10.1016/0550-3213(77)90252-8}{{\em Nucl. Phys.}
  {\bfseries B130} (1977) 505}.
[,329(1977)].
%%CITATION = NUPHA,B130,505;%%.

\bibitem{Vachaspati:1991dz}
T.~Vachaspati and A.~Achucarro, ``{Semilocal cosmic strings},''
  \href{http://dx.doi.org/10.1103/PhysRevD.44.3067}{{\em Phys. Rev. D}
  {\bfseries 44} (1991) 3067--3071}.

\bibitem{Hindmarsh:1991jq}
M.~Hindmarsh, ``{Existence and stability of semilocal strings},''
  \href{http://dx.doi.org/10.1103/PhysRevLett.68.1263}{{\em Phys. Rev. Lett.}
  {\bfseries 68} (1992) 1263--1266}.

\bibitem{Vachaspati:1992fi}
T.~Vachaspati, ``{Vortex solutions in the Weinberg-Salam model},''
  \href{http://dx.doi.org/10.1103/PhysRevLett.68.1977}{{\em Phys. Rev. Lett.}
  {\bfseries 68} (1992) 1977--1980}. [Erratum: Phys.Rev.Lett. 69, 216 (1992)].

\bibitem{Vachaspati:1992jk}
T.~Vachaspati, ``{Electroweak strings},''
  \href{http://dx.doi.org/10.1016/0550-3213(93)90189-V}{{\em Nucl. Phys. B}
  {\bfseries 397} (1993) 648--671}.

\bibitem{Achucarro:1999it}
A.~Achucarro and T.~Vachaspati, ``{Semilocal and electroweak strings},''
  \href{http://dx.doi.org/10.1016/S0370-1573(99)00103-9}{{\em Phys. Rept.}
  {\bfseries 327} (2000) 347--426},
  \href{http://arxiv.org/abs/hep-ph/9904229}{{\ttfamily arXiv:hep-ph/9904229}}.

\bibitem{ATLAS:2012yve}
{\bfseries ATLAS} Collaboration, G.~Aad {\em et~al.}, ``{Observation of a new
  particle in the search for the Standard Model Higgs boson with the ATLAS
  detector at the LHC},''
  \href{http://dx.doi.org/10.1016/j.physletb.2012.08.020}{{\em Phys. Lett. B}
  {\bfseries 716} (2012) 1--29},
  \href{http://arxiv.org/abs/1207.7214}{{\ttfamily arXiv:1207.7214 [hep-ex]}}.

\bibitem{CMS:2012qbp}
{\bfseries CMS} Collaboration, S.~Chatrchyan {\em et~al.}, ``{Observation of a
  New Boson at a Mass of 125 GeV with the CMS Experiment at the LHC},''
  \href{http://dx.doi.org/10.1016/j.physletb.2012.08.021}{{\em Phys. Lett. B}
  {\bfseries 716} (2012) 30--61},
  \href{http://arxiv.org/abs/1207.7235}{{\ttfamily arXiv:1207.7235 [hep-ex]}}.

\bibitem{Lee:1977qs}
B.~W. Lee and S.~Weinberg, ``{SU(3) x U(1) Gauge Theory of the Weak and
  Electromagnetic Interactions},''
  \href{http://dx.doi.org/10.1103/PhysRevLett.38.1237}{{\em Phys. Rev. Lett.}
  {\bfseries 38} (1977) 1237}.

\bibitem{Lee:1977tx}
B.~W. Lee and R.~E. Shrock, ``{An SU(3) x U(1) Theory of Weak and
  Electromagnetic Interactions},''
  \href{http://dx.doi.org/10.1103/PhysRevD.17.2410}{{\em Phys. Rev. D}
  {\bfseries 17} (1978) 2410}.

\bibitem{Barr:2008pn}
S.~M. Barr, ``{Doubly Lopsided Mass Matrices from Unitary Unification},''
  \href{http://dx.doi.org/10.1103/PhysRevD.78.075001}{{\em Phys. Rev. D}
  {\bfseries 78} (2008) 075001},
  \href{http://arxiv.org/abs/0804.1356}{{\ttfamily arXiv:0804.1356 [hep-ph]}}.

\bibitem{Chen:2023qxi}
N.~Chen, Y.-n. Mao, and Z.~Teng, ``{The global $B-L$ symmetry in the
  flavor-unified ${\rm SU}(N)$ theories},''
  \href{http://arxiv.org/abs/2307.07921}{{\ttfamily arXiv:2307.07921
  [hep-ph]}}.

\bibitem{Chen:2024cht}
N.~Chen, Y.-n. Mao, and Z.~Teng, ``{The Standard Model quark/lepton masses and
  the Cabibbo-Kobayashi-Maskawa mixing in an SU(8) theory},''
  \href{http://dx.doi.org/10.1007/JHEP12(2024)137}{{\em JHEP} {\bfseries 12}
  (2024) 137}, \href{http://arxiv.org/abs/2402.10471}{{\ttfamily
  arXiv:2402.10471 [hep-ph]}}.

\bibitem{Chen:2024deo}
N.~Chen, Z.~Hou, Y.-n. Mao, and Z.~Teng, ``{The gauge coupling evolutions of an
  SU(8) theory with the maximally symmetry breaking pattern},''
  \href{http://dx.doi.org/10.1007/JHEP10(2024)149}{{\em JHEP} {\bfseries 10}
  (2024) 149}, \href{http://arxiv.org/abs/2406.09970}{{\ttfamily
  arXiv:2406.09970 [hep-ph]}}.

\bibitem{Chen:2024yhb}
N.~Chen, Z.~Chen, Z.~Hou, Z.~Teng, and B.~Wang, ``{Further study of the
  maximally symmetry breaking patterns in an SU(8) theory},''
  \href{http://dx.doi.org/10.1103/wsm5-xfzt}{{\em Phys. Rev. D} {\bfseries 111}
  no.~11, (2025) 115034}, \href{http://arxiv.org/abs/2409.03172}{{\ttfamily
  arXiv:2409.03172 [hep-ph]}}.

\bibitem{Chen:2025ezv}
N.~Chen, J.~Tian, and B.~Wang, ``{The non-maximal symmetry breaking patterns in
  the supersymmetric $\widehat {\mathfrak{s} \mathfrak{u}}(8)_{k_U =1}$
  theory},'' \href{http://arxiv.org/abs/2508.11160}{{\ttfamily arXiv:2508.11160
  [hep-ph]}}.

\bibitem{James:1992zp}
M.~James, L.~Perivolaropoulos, and T.~Vachaspati, ``{Stability of electroweak
  strings},'' \href{http://dx.doi.org/10.1103/PhysRevD.46.R5232}{{\em Phys.
  Rev. D} {\bfseries 46} (1992) R5232--R5235}.

\bibitem{James:1992wb}
M.~James, L.~Perivolaropoulos, and T.~Vachaspati, ``{Detailed stability
  analysis of electroweak strings},''
  \href{http://dx.doi.org/10.1016/0550-3213(93)90046-R}{{\em Nucl. Phys. B}
  {\bfseries 395} (1993) 534--546},
  \href{http://arxiv.org/abs/hep-ph/9212301}{{\ttfamily arXiv:hep-ph/9212301}}.

\bibitem{Kanda:2022xrz}
Y.~Kanda and N.~Maekawa, ``{Stability of nontopological string in
  supersymmetric SU(2){\texttimes}U(1) gauge theory},''
  \href{http://dx.doi.org/10.1142/S0217751X22502219}{{\em Int. J. Mod. Phys. A}
  {\bfseries 37} no.~35, (2022) 2250221},
  \href{http://arxiv.org/abs/2205.12638}{{\ttfamily arXiv:2205.12638
  [hep-ph]}}.

\bibitem{Kanda:2023yyz}
Y.~Kanda and N.~Maekawa, ``{Stability of the embedded string in the
  SU(N){\texttimes}U(1) Higgs model and its application},''
  \href{http://dx.doi.org/10.1103/PhysRevD.107.096007}{{\em Phys. Rev. D}
  {\bfseries 107} no.~9, (2023) 096007},
  \href{http://arxiv.org/abs/2303.09517}{{\ttfamily arXiv:2303.09517
  [hep-ph]}}.

\bibitem{Goodband:1995rt}
M.~Goodband and M.~Hindmarsh, ``{Bound states and instabilities of vortices},''
  \href{http://dx.doi.org/10.1103/PhysRevD.52.4621}{{\em Phys. Rev. D}
  {\bfseries 52} (1995) 4621--4632},
  \href{http://arxiv.org/abs/hep-ph/9503457}{{\ttfamily arXiv:hep-ph/9503457}}.

\bibitem{Goodband:1995he}
M.~Goodband and M.~Hindmarsh, ``{Instabilities of electroweak strings},''
  \href{http://dx.doi.org/10.1016/0370-2693(95)01198-Y}{{\em Phys. Lett. B}
  {\bfseries 363} (1995) 58--64},
  \href{http://arxiv.org/abs/hep-ph/9505357}{{\ttfamily arXiv:hep-ph/9505357}}.

\bibitem{Eto:2024xvc}
M.~Eto, Y.~Hamada, R.~Jinno, M.~Nitta, and M.~Yamada, ``{Neutrino zeromodes on
  electroweak strings in light of topological insulators},''
  \href{http://dx.doi.org/10.1007/JHEP06(2024)062}{{\em JHEP} {\bfseries 06}
  (2024) 062}, \href{http://arxiv.org/abs/2402.19417}{{\ttfamily
  arXiv:2402.19417 [hep-ph]}}.

\bibitem{Chen:2021haa}
N.~Chen, Y.~Liu, and Z.~Teng, ``{Axion model with the SU(6) unification},''
  \href{http://dx.doi.org/10.1103/PhysRevD.104.115011}{{\em Phys. Rev. D}
  {\bfseries 104} no.~11, (2021) 115011},
  \href{http://arxiv.org/abs/2106.00223}{{\ttfamily arXiv:2106.00223
  [hep-ph]}}.

\bibitem{Chen:2021zwn}
N.~Chen, Y.-n. Mao, and Z.~Teng, ``{Bottom quark and tau lepton masses in a toy
  $\textrm{SU}(6)$ model},''
  \href{http://dx.doi.org/10.1140/epjc/s10052-023-11387-0}{{\em Eur. Phys. J.
  C} {\bfseries 83} no.~3, (2023) 259},
  \href{http://arxiv.org/abs/2112.14509}{{\ttfamily arXiv:2112.14509
  [hep-ph]}}.

\bibitem{Dimopoulos:1980hn}
S.~Dimopoulos, S.~Raby, and L.~Susskind, ``{Light Composite Fermions},''
  \href{http://dx.doi.org/10.1016/0550-3213(80)90215-1}{{\em Nucl. Phys. B}
  {\bfseries 173} (1980) 208--228}.

\bibitem{Earnshaw:1993yu}
M.~A. Earnshaw and M.~James, ``{Stability of two doublet electroweak
  strings},'' \href{http://dx.doi.org/10.1103/PhysRevD.48.5818}{{\em Phys. Rev.
  D} {\bfseries 48} (1993) 5818--5826},
  \href{http://arxiv.org/abs/hep-ph/9308223}{{\ttfamily arXiv:hep-ph/9308223}}.

\bibitem{Ambjorn:1989sz}
J.~Ambjorn and P.~Olesen, ``{Electroweak Magnetism: Theory and Application},''
  \href{http://dx.doi.org/10.1142/S0217751X90001914}{{\em Int. J. Mod. Phys. A}
  {\bfseries 5} (1990) 4525--4558}.

\bibitem{Achucarro:1993bu}
A.~Achucarro, R.~Gregory, J.~A. Harvey, and K.~Kuijken, ``{Cinderella
  strings},'' \href{http://dx.doi.org/10.1103/PhysRevLett.72.3646}{{\em Phys.
  Rev. Lett.} {\bfseries 72} (1994) 3646--3649},
  \href{http://arxiv.org/abs/hep-th/9312034}{{\ttfamily arXiv:hep-th/9312034}}.

\bibitem{Perkins:1993qz}
W.~B. Perkins, ``{W condensation in electroweak strings},''
  \href{http://dx.doi.org/10.1103/PhysRevD.47.R5224}{{\em Phys. Rev. D}
  {\bfseries 47} (1993) R5224--R5227}.

\bibitem{BianZY}
Z.~Bian. Github, 2026.
\newblock \url{https://github.com/mronebian/Non-topological-string-code.git}.

\bibitem{Bogomolny:1975de}
E.~B. Bogomolny, ``{Stability of Classical Solutions},'' {\em Sov. J. Nucl.
  Phys.} {\bfseries 24} (1976) 449.

\bibitem{Chen:2024wcj}
N.~Chen, Z.~Hou, and Z.~Teng, ``{The unification in an $ \hat{su}{(8)}_{k_U=1}
  $ affine Lie algebra},''
  \href{http://dx.doi.org/10.1007/JHEP04(2025)048}{{\em JHEP} {\bfseries 04}
  (2025) 048}, \href{http://arxiv.org/abs/2411.12979}{{\ttfamily
  arXiv:2411.12979 [hep-ph]}}.

\bibitem{Dvali:1993sg}
G.~R. Dvali and G.~Senjanovic, ``{Topologically stable electroweak flux
  tubes},'' \href{http://dx.doi.org/10.1103/PhysRevLett.71.2376}{{\em Phys.
  Rev. Lett.} {\bfseries 71} (1993) 2376--2379},
  \href{http://arxiv.org/abs/hep-ph/9305278}{{\ttfamily arXiv:hep-ph/9305278}}.

\bibitem{Eto:2018tnk}
M.~Eto, M.~Kurachi, and M.~Nitta, ``{Non-Abelian strings and domain walls in
  two Higgs doublet models},''
  \href{http://dx.doi.org/10.1007/JHEP08(2018)195}{{\em JHEP} {\bfseries 08}
  (2018) 195}, \href{http://arxiv.org/abs/1805.07015}{{\ttfamily
  arXiv:1805.07015 [hep-ph]}}.

\bibitem{Eto:2021dca}
M.~Eto, Y.~Hamada, and M.~Nitta, ``{Stable Z-strings with topological
  polarization in two Higgs doublet model},''
  \href{http://dx.doi.org/10.1007/JHEP02(2022)099}{{\em JHEP} {\bfseries 02}
  (2022) 099}, \href{http://arxiv.org/abs/2111.13345}{{\ttfamily
  arXiv:2111.13345 [hep-ph]}}.

\bibitem{Ambjorn:1988tm}
J.~Ambjorn and P.~Olesen, ``{ON ELECTROWEAK MAGNETISM},''
  \href{http://dx.doi.org/10.1016/0550-3213(89)90004-7}{{\em Nucl. Phys. B}
  {\bfseries 315} (1989) 606--614}.

\bibitem{Ambjorn:1989bd}
J.~Ambjorn and P.~Olesen, ``{A Condensate Solution of the Electroweak Theory
  Which Interpolates Between the Broken and the Symmetric Phase},''
  \href{http://dx.doi.org/10.1016/0550-3213(90)90307-Y}{{\em Nucl. Phys. B}
  {\bfseries 330} (1990) 193--204}.

\bibitem{Bimonte:1994ik}
G.~Bimonte and G.~Lozano, ``{Z flux line lattices and selfdual equations in the
  Standard Model},'' \href{http://dx.doi.org/10.1103/PhysRevD.50.R6046}{{\em
  Phys. Rev. D} {\bfseries 50} (1994) R6046--R6049},
  \href{http://arxiv.org/abs/hep-th/9403128}{{\ttfamily arXiv:hep-th/9403128}}.

\end{thebibliography}

\providecommand{\href}[2]{#2}\begingroup\raggedright\endgroup

\end{document}